\documentclass[twocolumn]{aastex631}

\newcommand{\orcid}[1]{\href{https://orcid.org/#1}
{\includegraphics[width=8pt]{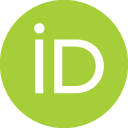}}}

\usepackage{amsmath}

\submitjournal{AJ}

\begin{document}

\title{A metal-poor atmosphere with a hot interior for a young sub-Neptune progenitor: \\
JWST/NIRSpec transmission spectrum of V1298 Tau b}

\correspondingauthor{Saugata Barat}
\email{saugatabarat500@gmail.com}

\author[0009-0000-6113-0157]{Saugata Barat}

\affiliation{Anton Pannekoek Institute for Astronomy, University of Amsterdam,
Science Park 904, 1098 XH,
Amsterdam, the Netherlands}

\affiliation{Kavli Institute for Astrophysics and Space Research, Massachusetts Institute of Technology, Cambridge, MA 02139, USA}

\affiliation{University of Southern Queensland, West St, Darling Heights, Toowoomba,
Queensland, 4350, Australia}

\author[0000-0002-0875-8401]{Jean-Michel D\'esert }
\affiliation{Anton Pannekoek Institute for Astronomy, University of Amsterdam,
Science Park 904, 1098 XH,
Amsterdam, the Netherlands}

\author[0000-0003-1622-1302]{Sagnick Mukherjee}
\affiliation{Department of Astronomy and Astrophysics, 
University of California, 
Santa Cruz, 95064, CA, USA}
\affiliation{Department of Physics and Astronomy, Johns Hopkins University, Baltimore, MD, USA}

\author[0000-0002-8515-7204]{Jayesh M. Goyal}
\affiliation{School of Earth and Planetary Sciences (SEPS), National Institute of Science Education and Research (NISER), Jatani, India}

\author[0000-0002-6215-5425]{Qiao Xue}
\affiliation{Department of Astronomy \& Astrophysics,
University of Chicago,
5640 S. Ellis Avenue,
Chicago, IL 60637,
USA}

\author[0000-0003-3800-7518]{Yui Kawashima}
\affiliation{Department of Astronomy, Graduate School of Science, Kyoto University, Kitashirakawa Oiwake-cho, Sakyo-ku, Kyoto 606-8502, Japan}
\affiliation{Frontier Research Institute for Interdisciplinary Sciences, Tohoku University, 6-3 Aramaki aza Aoba, Aoba-ku, Sendai, Miyagi 980-8578, Japan}
\affiliation{Department of Geophysics, Graduate School of Science, Tohoku University, 6-3 Aramaki aza Aoba, Aoba-ku, Sendai, Miyagi 980-8578, Japan}

\affiliation{Institute of Space and Astronautical Science, Japan Aerospace Exploration Agency, 3-1-1 Yoshinodai, Chuo-ku, Sagamihara, Kanagawa 252-5210, Japan}
\affiliation{Cluster for Pioneering Research, RIKEN, 2-1 Hirosawa, Wako, Saitama 351-0198, Japan}

\author[0000-0001-9504-3174]{Allona Vazan}
\affiliation{Astrophysics Reseach Center (ARCO), Department of Natural Sciences, The Open University of
Israel, Ra’anana, 43107, Israel}

\author[0000-0001-6315-7118]{William Misener}
\affiliation{Department of Earth, Planetary, and Space Sciences, The University of California, Los Angeles, 595 Charles E. Young Drive East, Los Angeles, CA 90095, USA}
\affiliation{Earth and Planets Laboratory, Carnegie Institution for Science, 5241 Broad Branch Road NW, Washington, DC 20015, USA}

\author[0000-0002-0298-8089]{Hilke E. Schlichting}
\affiliation{Department of Earth, Planetary, and Space Sciences, The University of California, Los Angeles, 595 Charles E. Young Drive East, Los Angeles, CA 90095, USA}

\author[0000-0002-9843-4354]{Jonathan J. Fortney}
\affiliation{Department of Astronomy and Astrophysics, 
University of California, 
Santa Cruz, 95064, CA, USA}

\author[0000-0003-4733-6532]{Jacob L. Bean}
\affiliation{Department of Astronomy \& Astrophysics,
University of Chicago,
5640 S. Ellis Avenue,
Chicago, IL 60637,
USA}

\author[0009-0000-7698-7057]{Swaroop Avarsekar}
\affiliation{School of Earth and Planetary Sciences (SEPS), National Institute of Science Education and Research (NISER), Jatani, India}

\author[0000-0003-4155-8513]{Gregory W. Henry}
\affiliation{Tennessee State University (Retired), Nashville, TN 37216 USA}

\author{Robin Baeyens}
\affiliation{Anton Pannekoek Institute for Astronomy, University of Amsterdam,
Science Park 904, 1098 XH,
Amsterdam, the Netherlands}

\author{Michael R. Line}
\affiliation{School of Earth and Space Exploration, 
Arizona State University, 
Tempe, AZ 85287, Arizona, USA}

\author[0000-0002-4881-3620]{John H. Livingston}
\affiliation{Astrobiology Center, 2-21-1 Osawa, Mitaka, Tokyo 181-8588,
Japan}
\affiliation{National Astronomical Observatory of Japan, 2-21-1 Osawa, Mitaka, Tokyo 181-8588, Japan}
\affiliation{Department of Astronomical Science, The Graduate University for Advanced Studies, SOKENDAI, 2-21-1 Osawa, Mitaka, Tokyo 181-8588, Japan}

\author[0000-0001-6534-6246]{Trevor David}
\affiliation{Center for Computational Astrophysics, Flatiron Institute, New
York, NY 10010, USA.}

\author[0000-0003-0967-2893]{Erik A. Petigura}
\affiliation{Department of Physics \& Astronomy, University of California
Los Angeles, Los Angeles, CA, 90095, USA}

\author[0000-0002-3522-5846]{James T. Sikora}
\affiliation{Lowell Observatory, 1400 W Mars Hill Road, Flagstaff, AZ, 86001, USA}

\author{Hinna Shivkumar}
\affiliation{Anton Pannekoek Institute for Astronomy, University of Amsterdam,
Science Park 904, 1098 XH,
Amsterdam, the Netherlands}

\author[0000-0002-9464-8101]{Adina D. Feinstein}
\altaffiliation{NHFP Sagan Fellow}
\affiliation{Laboratory for Atmospheric and Space Physics, University of Colorado Boulder, UCB 600, Boulder, CO 80309}
\affiliation{Department of Physics and Astronomy, Michigan State University, East Lansing, MI 48824, USA}

\author[0000-0002-9584-6476]{Antonija Oklop{\v{c}}i\'c}
\affiliation{Anton Pannekoek Institute for Astronomy, University of Amsterdam,
Science Park 904, 1098 XH,
Amsterdam, the Netherlands}

\begin{abstract}

We present the JWST/NIRSpec G395H transmission spectrum of the young (10–30 Myr) transiting planet V1298 Tau b (9.85$\pm$0.35~R$_{\oplus}$, T$_{eq}=670~K$). Combined HST and JWST observations reveal a haze-free, H/He-dominated atmosphere with a large scale height ($\sim$1500~km), allowing detection of  CO$_2$ (35$\sigma$), H$_2$O (30$\sigma$), CO (10$\sigma$), CH$_4$ (6$\sigma$), SO$_2$ (4$\sigma$) and OCS (3.5$\sigma$). Our observations probe several scale heights ($\sim$4.4 in the CO$_2$ 4.3$\mu$m band and $\sim$3 in the 2.7$\mu$m water band). The planet’s mass, inferred from atmospheric scale height using free retrieval and grid modeling, is 12$\pm$1~M$_{\oplus}$ and 15$\pm$1.7~M$_{\oplus}$, respectively, which is significantly lower than previous radial velocity estimates and confirms it as `gas-dwarf' sub-Neptune progenitor. We find an atmospheric metallicity (logZ=0.6$^{+0.4}_{-0.6}\times$solar) and sub-solar C/O ratio (0.22$^{+0.06}_{-0.05}$). The atmospheric metallicity is low compared to mature sub-Neptunes by an order of magnitude. The CH$_4$ abundance ([CH$_{4}$]=-6.2$^{+0.3}_{-0.5}$) is $\sim$7$\sigma$ lower than equilibrium chemistry prediction. To adjust for the low methane abundance,  the self-consistent grids favor a high internal temperature ($\sim$500~K) and vertical mixing (K$_{zz}\sim10^{7}-10^{8}$~cm$^2$/s). These internal temperatures are inconsistent with predictions from evolutionary models, which expect $\sim$ 100–200~K at the current system age. We estimate a gas-to-core mass fraction between 0.1–8\%, with a core mass of 11-12~M$_{\oplus}$, consistent with in-situ gas-dwarf formation. A deep atmospheric metallicity gradient may explain both the high internal temperature and low observable metallicity. Over time, mass loss from such an atmosphere could enhance its metallicity, potentially reconciling V1298 Tau b with mature sub-Neptunes.

\end{abstract}


\section{Introduction}

In recent years, a number of young transiting planets have been discovered by \texttt{Kepler}, \texttt{K2} and \texttt{TESS} \citep[e.g see][]{david2016,mann2016,newton2019,plavchan2020,mann2022}. Many of these young transiting planets appear to have super-Neptunian or even Jovian sizes and occupy a rare part of the period-radius plane for exoplanets. However, based on early evolutionary models, these planets are more likely to be sub-Neptune/super-Earth progenitors rather than gas giants, undergoing thermal cooling and contraction \citep{Lopez_2014,owen_2020,kubyshkina2020}. Comparing the size distribution of the young transiting planet population, \citet{vach2024} also concluded that these young planets are most likely to be sub-Neptunes/super-Earths undergoing early evolution.

In the core-accretion model \citep{pollack1996}, sub-Neptunes and super-Earths can be formed similarly to gas giants, i.e. by the formation of a rocky core followed by accretion of gas from the disk. However, these planets do not enter a runaway gas accretion and can end up with a gas-to-core ratio (GCR) of the order of 10\% after formation \citep{lee2014}. This mechanism leads to the formation of the H/He rich gas-dwarfs \citep{rogers2025}. These planets are expected to lose significant mass after formation driven by photoevaporation \citep{owen2013,owen2019} and/or core-powered mass loss \citep{ginzburg2018,owen_schlichting2024}. Recent evolutionary models suggest that for these low-mass planets, early atmospheric mass loss could result in significant evolution of its atmospheric composition \citep{Malsky2020}. However, the composition of gas giants is not expected to be significantly affected by early mass loss mechanisms \citep{louca2023}. 

Atmospheric composition evolution due to early mass loss is expected to be even more significant for primordial envelopes with metallicity gradients, i.e., gradually decreasing metallicity in the atmosphere \citep{fortney2013}. Metallicity gradients have been reported for solar system giant planets \citep{wahl2017,vazan2020,2mankovich2021} and could be due to formation through pebble accretion \citep{ormel2021,bloot2023}, or chemical equilibrium between H/He dominated atmosphere and silicate dominated magma \citep{misener2022,Misener2023}. 
Another mechanism associated with primordial composition gradients is rainout from the outer parts of the gradient as the planet cools. Silicate rainout at early evolution has significant consequences on the thermal evolution and mass loss of sub-Neptunes \citep{2vazan2023,vazan2024}.

Atmospheric characterization of young transiting planets offers a unique opportunity to test these predictions from early evolutionary theories. The V1298 Tau system is one of the youngest (10-30~Myr old) multi-planet systems \citep{david2019,mascareno_2021} located in the moving group 29 in the foreground of Taurus-Auriga \citep{oh17,luhman18}. These studies consider the age of the star to be the same as that of the moving group it is a member of, and constrains the age of the moving group by fitting the isochrone.  Other suggest that the star is younger, with an age of $\sim$ 10~Myr \citep{finoceti2023,maggio2022}. However, these studies estimate the age based on comparing the luminosity and temperature of V1298 Tau with evolutionary models, therefore do not take into account observations of the rest of the group members. In this paper we adopt the age reported by fitting isochrones to Group 29 (20$\pm$10~Myr).  The host star is a weak-lined T-Tauri star with a rotation period of $\sim$2.8 days. It has 4 transiting planets, the inner three found in a near 2:3:6 mean-motion resonance \citep{david2019}, with radii between 5-10~R$_{\oplus}$. The system is well aligned \citep{johnson2022,gaidos2022}. Atmospheric mass loss has not been detected for V1298 Tau b and c, with a tentative detection for V1298 Tau d \citep{feinstein2021, vissapragada_21,alam2024}.

This paper focuses on V1298 Tau b (9.8$\pm$0.355~$R_{\oplus}$; 24.139$\pm$0.001 day period, $\sim$670~K equilibrium temperature assuming full recirculation and 0 albedo \citep{david2019b}). Mass measurements for young transiting planets have been challenging because of the RV jitter from stellar activity \citep{bremms2019, tran2024}. Initial RV measurements of the V1298 Tau system reported a Jovian mass ($0.64 \pm 0.19 M_J$, \citet{mascareno_2021}) for V1298 Tau b. Subsequent RV follow-up work reported the mass upper limit for V1298 Tau b; $<$159~M$_{\oplus}$ \citep{sikora2023} and $< 100 M_{\oplus}$ \citep{finoceti2023}. The challenges of modeling RV from active stars, such as V1298 Tau, and the biases it can introduce in RV mass measurements have been shown in \citet{blunt2023}.

The HST/WFC3 transmission spectrum of V1298 Tau b \citep{barat2023}, revealed a large water absorption feature (400~ppm) at 1.4$\mu$m, indicating an extended and haze-free atmosphere. The HST/WFC3 transmission spectrum was used to estimate an upper limit of 23$M_\oplus$ at 3$\sigma$ confidence level from the large scale height observed for this planet. The HST/WFC3 transmission spectrum ruled out the RV mass estimate \citep{mascareno_2021} at 5$\sigma$ confidence level. Ongoing monitoring of transit timing variations (TTVs) in the system yields a mass in agreement with the upper limit from HST (Livingston et al. in prep.). Atmospheric models showed that the atmospheric metallicity is low ($\sim$ solar). 

A low abundance of methane has been reported from the HST observations of V1298 Tau b \citep{barat2023}. Given its equilibrium temperature (670~K), it is expected to be rich in methane based on equilibrium chemistry \citep{fortney_2020}. JWST observations WASP-107b, another low density super-Neptune mass planet with a similar equilibrium temperature (750~K) has also revealed a significantly methane depleted atmosphere \citep{welbanks2024,sing2024}. This lack of methane in warm transiting planets has been previously discussed from Spitzer photometry \citep{Baxter2021}, and hints towards strong vertical mixing and high internal temperatures ($>$400~K) such that methane poor gas from the deep atmosphere is dredged up to the observable atmosphere. Therefore, methane acts as an indirect probe of the thermal conditions of the deep atmosphere (`methane-thermometer'). Such high internal temperatures are inconsistent with formation and evolution models of Neptune or lower mass planets and could require heating mechanisms, such as tidal heating. Tidal heating has been considered as a potential explanation for the high internal temperature inferred for WASP-107b \citep{sing2024,welbanks2024}.

In this paper, we present observations of V1298 Tau b using JWST/NIRSpec G395H \citep{birkman2022} and report the resulting near-infrared (NIR) transmission spectrum. The paper is summarized as follows. In Section \ref{observations} we describe the JWST and ground-based observations of V1298 tau b. In Section \ref{section:data analysis}, we discuss the reduction of JWST data and the analysis of light curves. In Section \ref{section:results}, we present the transmission spectrum and atmospheric modeling. In Section \ref{section:discussion}, we discuss our findings and interpret our results in the context of planet formation and early evolution.

\section{Observations} \label{observations}

\subsection{JWST Observations}
We observed a primary transit of V1298 Tau b as a part of the JWST Cycle 1 GO program 2149 (PI: D\'esert) using NIRSpec Bright Object Time-Series Spectroscopy (BOTS) G395H/SUB2048, on 12th February 2023. We used 4 groups per integration resulting in 10250 integrations of 4.51 seconds each, resulting in 12.8 hours of science observations. The total visit lasted 14.7 hours with 2 hours of overheads. None of the detector pixels exceeded 30$\%$ of the saturation limit, ensuring operations within the linear regime of the detector. The raw data used for this analysis is provided in \dataset[doi:10.17909/kjg5-8t66]{https://doi.org/10.17909/kjg5-8t66}.

\subsection{Ground based long-term photometric monitoring of V1298 Tau} \label{ground based data description}

 We acquired 509 observations of V1298 Tau in the Cousins $R$ band with the
TSU Celestron 14-inch (C14) Automated Imaging Telescope (AIT) from late in
the 2019-20 observing season through the 2023-24 season.  Details of the instruments used and data analysis are provided in Appendix \ref{v1298_photometric_monitoring}.

We show the differential R-band magnitudes for season 2 (coincident with HST visit) and season 5 (coincident with JWST visit) in Figure \ref{fig:v1298_photometry}. The optical R-band shows significant variability both on nightly and yearly timescales. We performed a periodogram analysis on the individual observing seasons using 
the method of \citet{vanikek1971}.  The mean of the rotation periods of all observing seasons is $2.898\pm0.011$ and is consistent with the rotation period inferred from the TESS and K2 observations \citep{david2019b,Feinstein2022}. We also find a peak-to-peak amplitude of $\sim$4\% during season 2 and 5, which is also consistent with the variability amplitude reported from the K2 and TESS observations.

\section{Data analysis} \label{section:data analysis} 

\subsection{JWST data reduction} \label{data reduction}

We started our analysis from the raw uncalibrated (*uncal.fits) files downloaded from the MAST archive. We used two independent data reduction pipelines: \texttt{Eureka!} \citep{eureka} and \texttt{SPARTA} \citep{kempton2023}. \texttt{Eureka!} has been applied to several JWST time series observations using NIRSpec/G395H \citep[e.g.][]{May2023,benneke2024}.  

In the first two stages of data reduction \texttt{Eureka!} uses the \texttt{jwst} pipeline to convert ramps-to-slopes, apply column-by-column 1/f correction at the group stage, remove cosmic rays. In the third stage, \texttt{Eureka!} uses custom tools to perform an optimal extraction \citep{horne1986}. First, it aligns the curved trace across the detector and again applies a column-by-column background subtraction. It constructs a median of all frames to calculate the weights requires for optimal extraction. It performs column-by-column optimal extraction to derive 1D stellar spectra from our observations. We only use \texttt{Eureka!} till Stage 3. The detailed steps of the data reduction and conversion from raw images to 1D stellar spectra are given in Appendix \ref{eureka}.

\texttt{SPARTA} uses custom software from Stage 1. It is similar to \texttt{Eureka!} in its reduction procedure. It uses group level row-by-row and column-by-column background subtraction to account for 1/f noise.  It applies the same background subtraction at the integration level (Stage 3).  The description of the data reduction from \texttt{SPARTA} is described in Appendix \ref{sparta}.

\subsection{JWST Light curve analysis} \label{subsec:jwst_light curve}

\begin{figure*}
    \centering
    \includegraphics[width=1\textwidth]{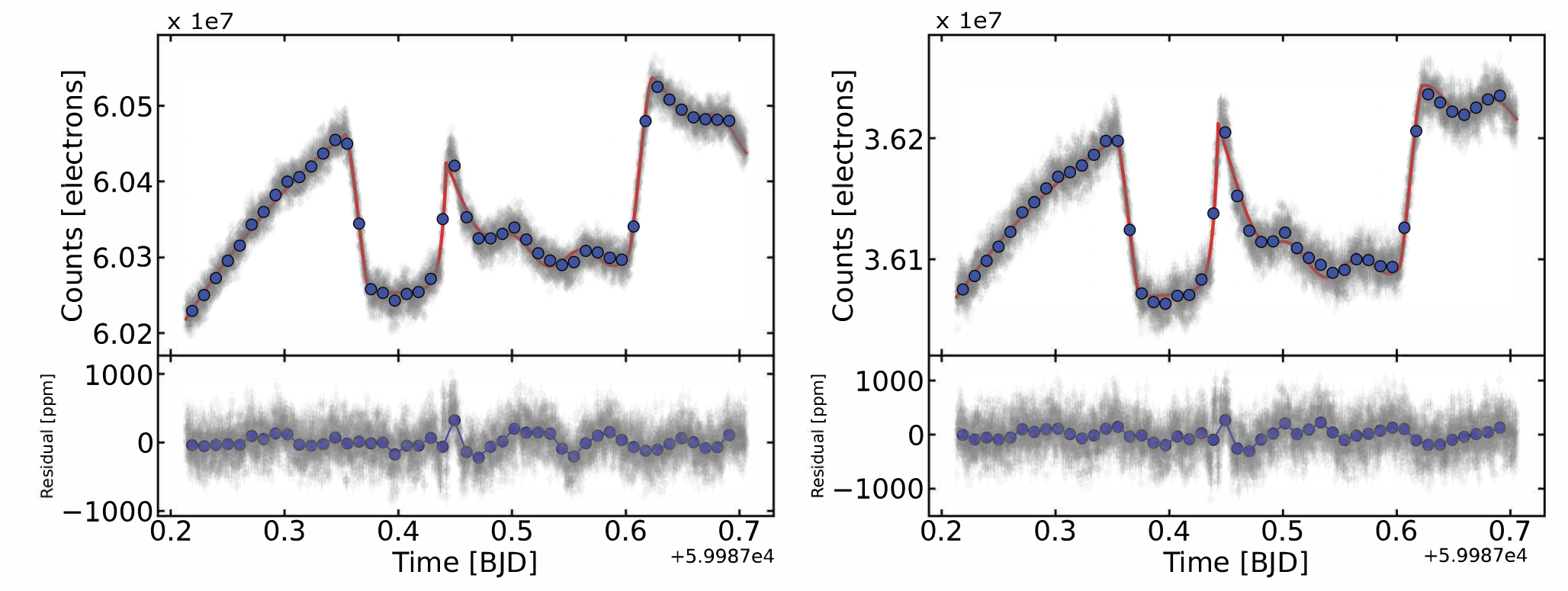}
    \caption{Extracted white light curve from the two NIRSpec G395H detectors, NRS1 (left) and NRS2 (right) from the \texttt{Eureka!} data reduction (Appendix \ref{eureka}). The upper panel shows the observed counts for each integration (grey points) and the best fit empirical model (red solid line, See Appendix \ref{subsec:lightcurve_analysis}). The transit light curves show a large flare in transit followed by post-flare oscillations. The baseline also appears curved which is due to the rotational variability of the young T-Tauri host star. The empirical model (Eqn \ref{eq:eq1}) used to fit the light curves consists of a rotational variability term (quadratic polynomial in time), double exponential flare model, a sinusoid for the post flare oscillations and the planet transit modeled using \texttt{batman} \citep{batman}. We use an MCMC framework to fit the empirical model to the observed light curve and calculate the best-fit model using median values from the posterior distributions of the fitted parameters. The lower panel shows the residuals obtained by subtracting the best-fit model from the observed white light curve. The residuals show significant structure indicating that the empirical model fit does not sufficiently account for all the astrophysical systematics.   }
    \label{fig:white_lc}
\end{figure*}

In Figure \ref{fig:white_lc} we show the white transit light curve of V1298 Tau b extracted from the Eureka! data reduction. We see low frequency variability consistent with rotational variability of the star and a high frequency signal which resembles a stellar flare from it's rapid rise and gradual decay temporal behaviour. We also see post-flare oscillations in the transit light curve which continues even after egress. We model the white light curves and the spectroscopic light curves by fitting an empirical model. We follow-up this analysis with a data-driven method to detrend the long term and short term variability and re-fit the spectroscopic light curves. Using this approach we can reach within $\sim$5\% of the expected photon noise for all spectroscopic bins, except eight bins. We discuss the nature of the light curves, the empirical model used for the fit and the data-driven approach in detail in Appendix \ref{subsec:lightcurve_analysis}.

\begin{figure*}
    \centering
    \includegraphics[width=1\linewidth]{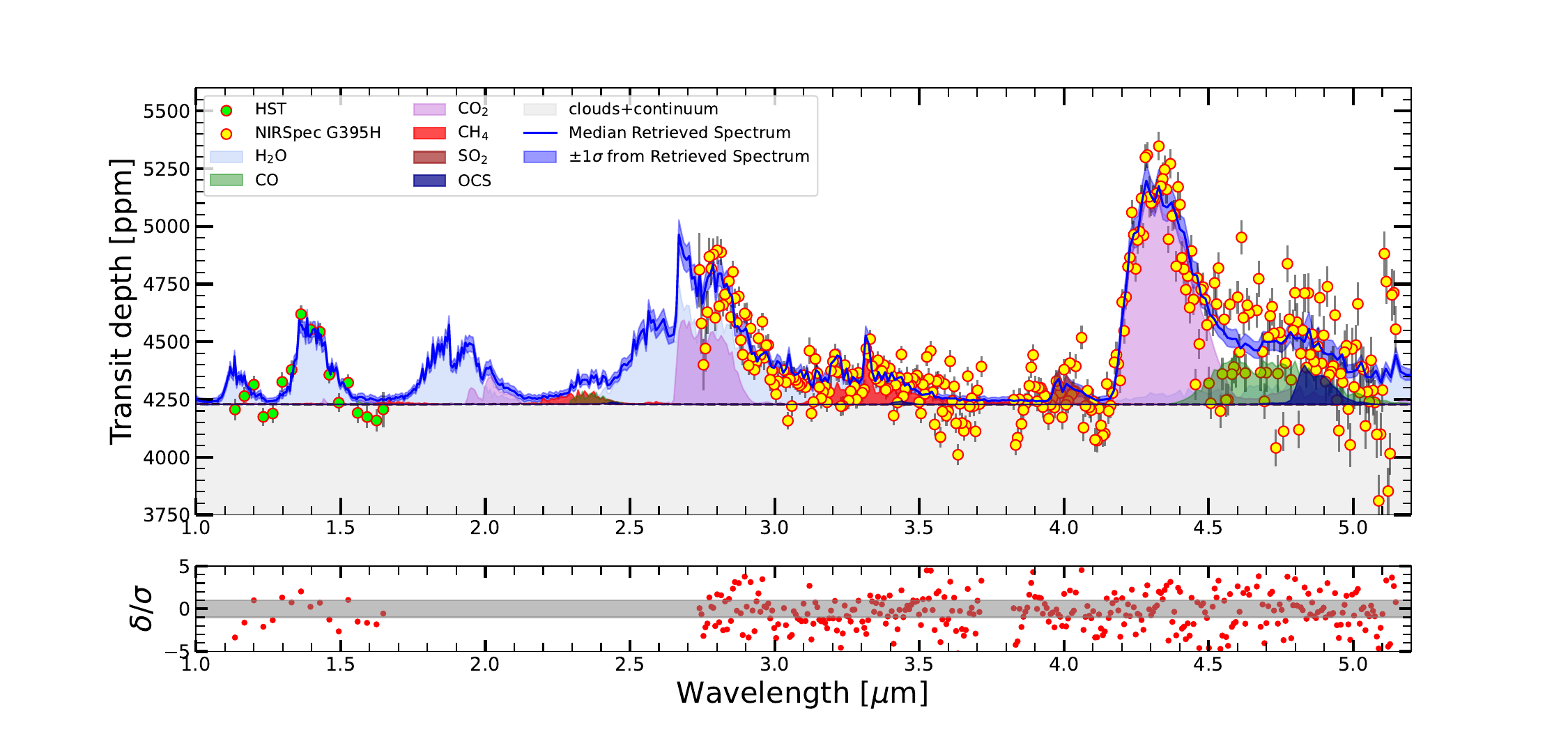}
    \caption{Transmission spectrum of V1298 Tau b observed using JWST/NIRSpec G395H (yellow points), and using HST/WFC3 \citep{barat2023} (green points),  with $1\sigma$ uncertainties, along with best fit model (blue solid) from the free chemistry PICASO retrieval (see Section \ref{spectrum} and Appendix \ref{appendix: free retrieval}). The JWST spectra and their $1~\sigma$ errorbars have been derived from the \texttt{Eureka!} data reduction (Appendix \ref{eureka}), using data-driven flare corrected light curve fitting methods (Appendix \ref{data-driven detrending}). The HST spectra has been offset by a 400~ppm to adjust for the difference in brightness level of the star between the two epochs estimated from ground based photometric long term monitoring (Appendix \ref{v1298_photometric_monitoring}). The transmission spectrum shows a mostly clear atmosphere with a large scale height ($\sim$1500km). The contribution from H$_2$O (30$\sigma$), CH$_4$ (6$\sigma$), CO$_2$ (35$\sigma$), CO (10$\sigma$) and the continuum opacity (grey clouds) are apparent from the PICASO free retrieval and are shown with the colored shaded regions.  In the bottom panel we show the residuals (scaled with the errorbar on each bin) between the combined JWST and HST observations with the best fit PICASO free chemistry model. In the HST bandpass, NRS1 and NRS2 the RMS residual 47 ppm, 55 ppm and 78 ppm respectively. The grey shaded region in the lower panel shows the 1$\sigma$ confidence region.}
    \label{fig:gas contribution}
\end{figure*}

 In Figure \ref{fig:detrended_lc} we show stacked detrended spectroscopic light curves obtained using the data-driven technique. It shows that the low frequency variability signal has been removed from the spectroscopic light curves. However, four horizontal streaks are seen across the stacked detrended light curves, which we identify as four stellar lines where the data-driven de-trending method does not work due to different temporal behaviour (flare shape) compared to the rest of the continuum. We remove these channel, which correspond to 8 spectral bins from the rest of the analysis. We further discuss the detrended light curves and the outliers in detail in Appendix \ref{data-driven detrending}. For the rest of the analysis we have used the transmission spectrum extracted using the Eureka!, unless otherwise specified. The transmission spectrum we derive from the data-driven spectroscopic light curve fits is shown in Figure \ref{fig:gas contribution}.

\begin{figure*}
    \centering
    \includegraphics[width=\linewidth]{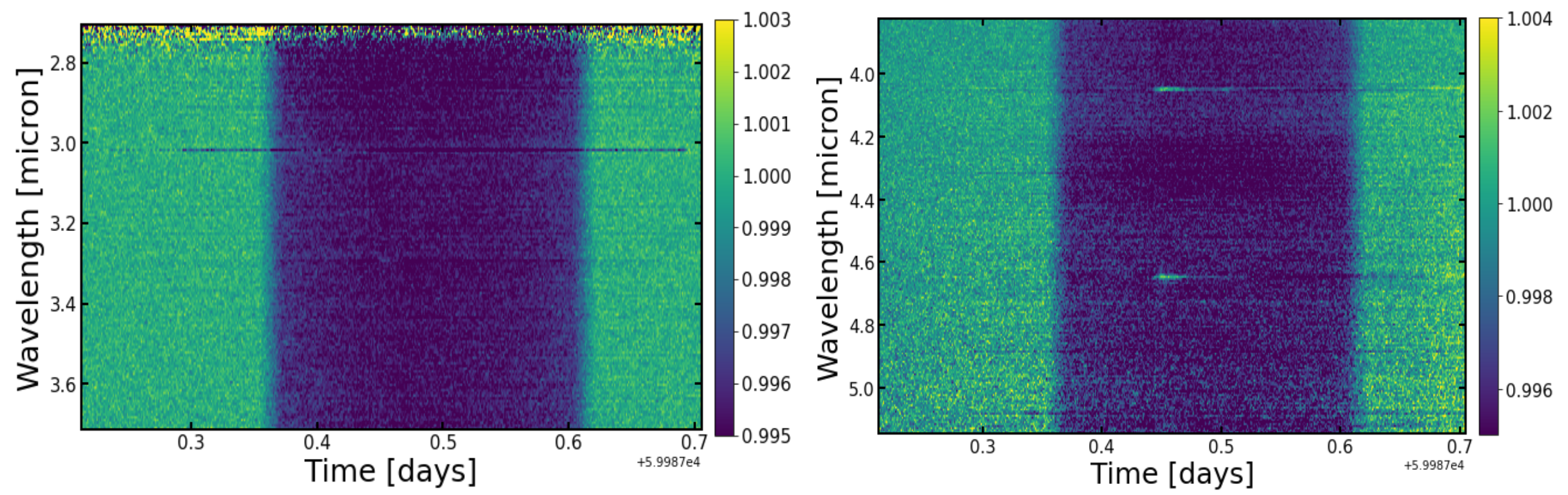}
    \caption{Stacked detrended spectroscopic light curves (left NRS1, right NRS2) derived using data driven detrending approach described in Appendix \ref{subsec:lightcurve_analysis}. First, the spectroscopic light curves were fitted using an empirical model (Eqn \ref{eq:eq1}). Following this, the second order polynomial coefficients derived from the empirical model fit were smoothed in wavelength using a gaussian kernel (Figure \ref{figure:polynomial_spectrum_correlation}), and the smoothed coefficients were used to create a stellar variability model for each spectroscopic bin. The spectroscopic light curves were detrended using this variability model. The flare profile is derived from the best-fit flare parameters from the white light curve fit and the oscillation profiles is calculated by removing the stellar baseline model, transit and flare profile from the white light curve (Figure \ref{fig:flare_profile}). The detrending flare and oscillation profiles were scaled for each spectroscopic light curve, and the scaling factors were included as fitting parameters in the second iteration of the spectroscopic light curve fits (Section \ref{data-driven detrending}) to account for the wavelength dependence of the flare and oscillation amplitude. We can identify four outliers by eye on this diagram. These bins can be seen in this figure as horizontal stripes. We identify these bins at 3.03, 3.29, 4.04 and 4.65 microns and remove them from the subsequent analysis. From visual inspection of the light curves at these bins we find that the temporal shape of the flare/oscillations are different in these bins compared to the median profile used for detrending. Therefore, the data driven detrending do not work well for these channels.  The transit and the ingress/egress are visible by eye on the stacked detrended light curves. In the right panel, a darker patch is visible between 4.2 and 4.4 micron, which is due to CO$_2$ absorption. }
    \label{fig:detrended_lc}
\end{figure*}

\begin{figure*}
    \centering
    \includegraphics[width=0.5\linewidth]{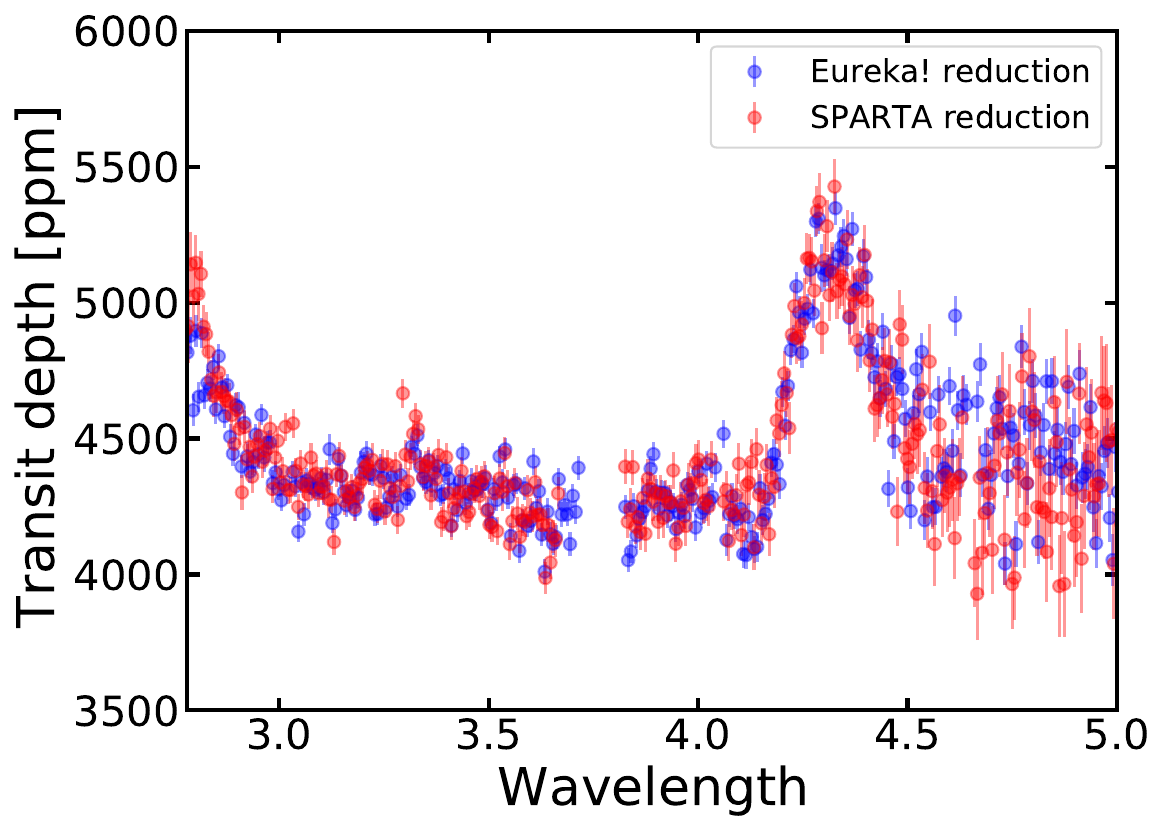}
    \caption{Comparison between transmission spectrum of V1298 Tau b derived from \texttt{Eureka!} (Appendix \ref{eureka}) and \texttt{SPARTA} (Appendix \ref{sparta}) data reduction. The transmission spectra shown here have been derived using the data-driven light curve fitting method as discussed in Appendix \ref{data-driven detrending} applied independently to the \texttt{Eureka!} and \texttt{SPARTA} data reductions. The transmission spectra from both reductions are consistent with each other and show similar molecular absorption features. In NRS1 both data reductions show a similar level of scatter, however, the SPARTA transmission spectrum shows higher scatter in NRS2 compared to Eureka!. This is likely due to the slightly higher scatter we found in the NRS2 1D stellar spectra from SPARTA compared to Eureka! and this difference in scatter for the SPARTA reduction is also seen in the white light curve comparison (Figure \ref{fig:eureka_sparta_comparison}).  }
    \label{fig:eureka_sparta_spectra_comparison}
\end{figure*}

Using the same light curve analysis methodology outlined in Appendix \ref{subsec:lightcurve_analysis}, we analyze the light curves extracted from the SPARTA data reduction. A comparison between the Eureka! and SPARTA light curves is shown in Figure \ref{fig:eureka_sparta_comparison}. The white light curves show similar mean absolute deviation (MAD) in NRS1 (129 ppm and 133 ppm for Eureka and SPARTA respectively), but in NRS2 SPARTA shows slightly higher MAD (230 ppm) compared to Eureka! (190 ppm).  We derive the transmission spectrum from the \texttt{SPARTA} reduction (Appendix \ref{sparta}) using the data-driven detrending approach outlined in this section (Figure \ref{fig:eureka_sparta_spectra_comparison}). We find that the NRS1 spectra are consistent between both data reductions, however, for the NRS2 the SPARTA reduction shows slightly higher scatter at the red end of the spectrum (Figure \ref{fig:eureka_sparta_comparison}). This is due to the relatively high scatter in the SPARTA NRS2 light curves compared to Eureka! We have developed SPARTA as an independent data reduction pipeline for NIRSpec/G395H for this paper. However, it is still under development, which could be the reason for the higher scatter in NRS2. Further development of the code could improve the reduction of NRS2 data in the future However, it can be concluded that the light curves of the independent reductions show similar features and the spectra also show similar atmospheric absorption features.

\subsection{Combining JWST and HST spectra of V1298 Tau}

V1298 Tau b was observed during a primary transit using HST/WFC3 \citep{barat2023} which provides complementary wavelength coverage to the current JWST observations. Combining the HST observations with the current JWST observations is likely to be beneficial for constraining atmospheric properties because of the broader wavelength coverage.

However, transit radii from different epochs of active stars can differ because of the changing flux of the host star \citep{czesla2009,desert2011b}. The variation in transit depth due to flux variation of the star is given by Eqn \ref{eq:eq0}, where F$_2$ and F$_`$ refer to the flux level of the star at the two epochs. $\alpha$ is an empirical constant of the order of unity \citep{desert2011b}. It is equal to -1 if the brightness of the part of the photosphere not covered by spots is uniform (i.e., there are no faculae). The HST observations of V1298 Tau b were taken in October 2020 and the JWST observations in February 2023, separated by almost 3 years. Thus, it is important to assess the effects that could be introduced because of the flux variation of the star prior to combining the observations from two different epochs.

\begin{equation} \label{eq:eq0}
  \frac{\Delta(R_{p}/R_{*})}{R_{p}/R_{*} } \sim \alpha f_{\lambda} / 2 
    \end{equation}

\begin{equation} \label{eq:eq01}
    f_{\lambda} = \frac{F_{2}(\lambda)-F_{1}(\lambda)}{F_{1}(\lambda)}
\end{equation}

 The white light transit depth (best fit) from HST is 4150 ppm \citep{barat2023}, and JWST NRS2 is 4540 ppm. We find that the host star is brighter by 0.08 mag in the R-band on the night of the HST visit compared to the JWST visit. We calculate the expected change in transit depth due to this variation in the stellar brightness between the two epochs using the formalism outlined in \citet{desert2011b} and shown in Eqn \ref{eq:eq0} and Eq \ref{eq:eq01}.  We assumed a photosphere of 5000~K for V1298 Tau and spots with temperature of 4000~K \citep{biagini2024}.  We find that $\alpha$=-1.75 can explain the apparent difference in the white-light transit depth between the two epochs. A value of $\alpha$ lower than -1 implies that the part of the photosphere which is not covered by spots for V1298 Tau b is not uniform, there could be faculae. Interestingly, \citet{barat2023} had reported a bright spot crossing. 

We show the combined JWST and HST transmission spectra in Figure \ref{fig:spectra_full}, where the HST data have been offset by 400 ppm.

\section{Results} \label{section:results}

\subsection{Transmission spectrum of V1298 Tau b} \label{spectrum}

The JWST/NIRSpec G395H and the HST/WFC3 G141 transmission spectra of V1298 Tau b (Figure \ref{fig:gas contribution}) are consistent with each other. The combined transmission spectrum shows a relatively clear atmosphere with a large scale height ($\sim$1500~km) and a large water absorption feature around 2.8-3~$\mu$m. Around 4.3$\mu$m a large 1000~ppm CO$_2$ absorption feature can be seen. The absorption feature is so strong that it can already be seen in the detrended NRS2 spectroscopic light curves (Figure \ref{fig:detrended_lc}), even without fitting them.

We perform free chemistry atmospheric retrievals on the combined HST/WFC3 and NIRSpec/G395H spectrum using the open-source \texttt{PICASO} atmospheric model \citep{batalha19,Mukherjee22}. We include 12 molecular species, gray cloud opacity, planet mass, and offset between the JWST and HST spectrum as well as between the NRS1 and NRS2 detectors. We assume an isothermal 1D plane-parallel atmosphere. We discuss details, such as the priors adopted, details of the sampling method used and estimation of molecular detection significances in Appendix \ref{appendix: free retrieval}. Although optical monitoring predicts $\sim$400 ppm offset, we include an offset parameter between the HST and JWST data in the retrieval. The posterior distribution obtained with this retrieval setup is shown in Figure \ref{fig:v1298 free chemistry corner}. From the free retrieval we detect H$_2$O (30$\sigma$), CH$_4$ (6$\sigma$), CO$_2$ (35$\sigma$), CO (10$\sigma$), as well as tentative evidence for OCS (3.5$\sigma$) and SO$_2$ (4$\sigma$), while upper limits are obtained for the abundance of the rest of the gases (NH$_3$, H$_2$S, HCN, N$_2$ and C$_2$H$_2$) included in the free chemistry retrieval. 

Using the VMR from the free retrieval, we can calculate the metallicity and C/O ratio. For metallicity, we calculated the ratio between the number of atoms of all species heavier than H/He with the number of H and He atoms, and for the C/O ratio we calculated the ratio of C and O atoms from our models. We compare these values with solar abundances \citep{asplund_2009}. Using the retrieved molecular abundances, we constrain the atmospheric metallicity of the planet to be [M/H]= +0.6$^{+0.4}_{-0.6}$ solar. We find that C / O is 0.22$^{+0.06}_{-0.05}$ from the free retrieval. The retrieved metallicity is consistent with the solar metallicity reported in \citet{barat2023} within 1$\sigma$. We find log(CH$_4$)=$-6.22\pm0.42$, which is lower than the methane detectability limit for HST / WFC3 G141 \citep{fortney_2020}, which confirms the low methane abundance reported from the HST spectrum.  The contribution of each gas to the retrieved median spectrum is shown in Figure \ref{fig:gas contribution}. The free retrieval measures a mass of 12$\pm$1M$_{\oplus}$ for V1298 Tau b.

The isothermal free chemistry retrieval constrains the temperature of the planet at $T= 444 \pm 38$K. We note that this photospheric temperature is colder than the expected equilibrium temperature (670~K) assuming 0 albedo and full day-to-night side recirculation \citep{david2019b} by about $\sim$230~K.  We also detect the presence of a gray cloud deck in the atmosphere with a cloud opacity of log($\kappa_{\rm cld}$)= $-2.77\pm0.23$, where $\kappa_{\rm cld}$ is in cm$^2$/g. The free retrieval finds an offset of 405$\pm$15~ppm between the HST and JWST observations, but the offset between the two NIRSpec detectors is consistent with 0.

\begin{figure*}
    \centering
    \includegraphics[width=1\textwidth]{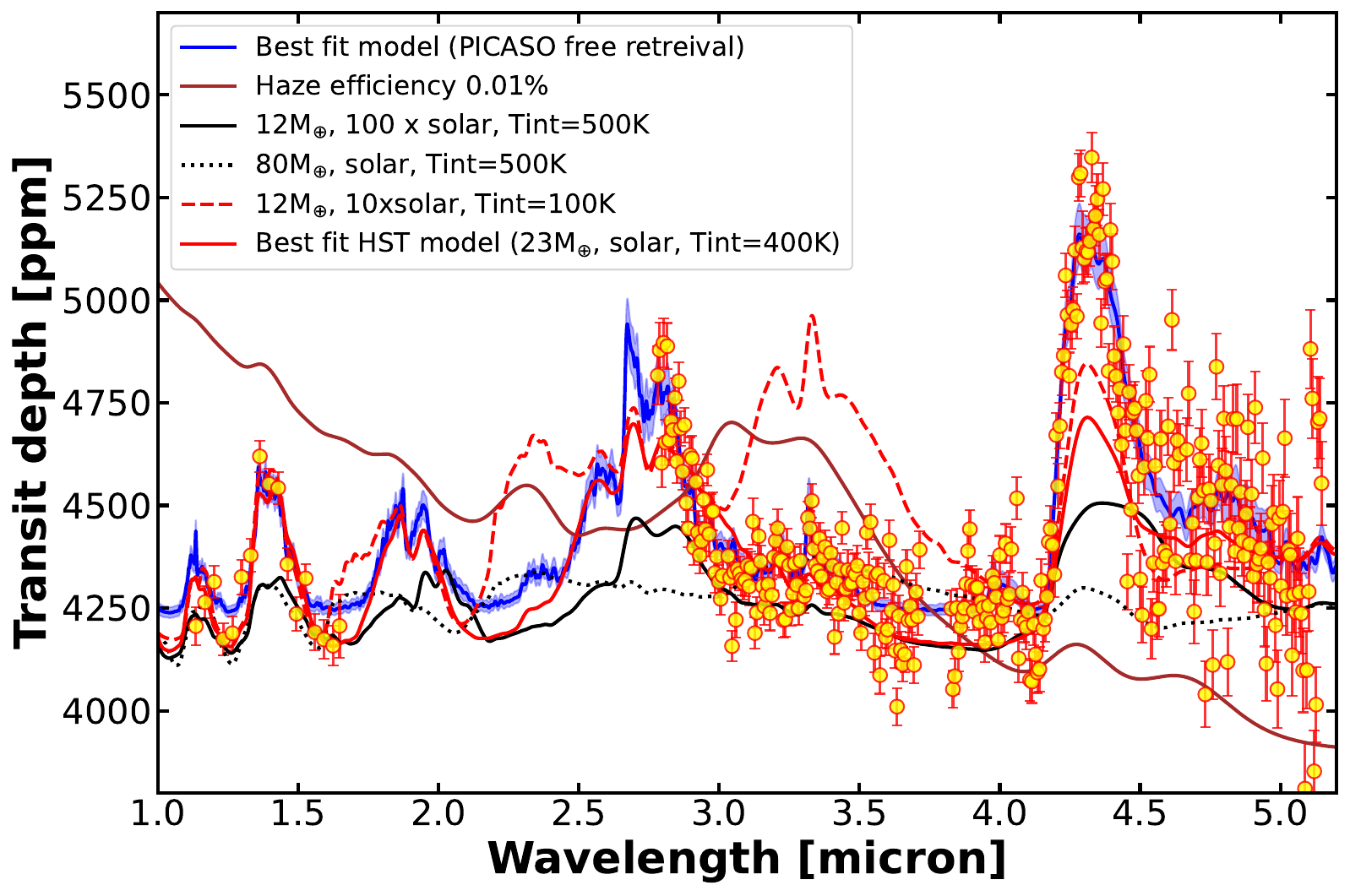}
    \caption{Transmission spectrum of V1298 Tau b observed with JWST/NIRSpec G395H and HST/WFC3 (Section \ref{spectrum}), with $1\sigma$ errobars, along with some models which were nominally expected based on the published RV mass \citep{mascareno_2021}, equilibrium temperature, predicted internal temperature from evolution models (Figure \ref{fig:interior_structure_model}) and the mass-metallicity relation for core-accretion formation\citep{thorngren2016}. The blue solid line shows the best fit model from the PICASO free retrieval (see Section \ref{spectrum} and Appendix \ref{appendix: free retrieval}) and the shaded region around it shows the 1$\sigma$ confidence interval. The black solid line shows a model with 12~M$_{\oplus}$, 100$\times$ solar (assumed from the mass-metallicity relation) and T$_{int}$=500~K which can ruled out at 23$\sigma$ confidence level. The model rejection significance is estimated from the difference in Chi-square values between the best fit model (blue) and the model being compared.  The black dotted line shows a solar metallicity, 80~M$_{\oplus}$ (2$\sigma$ mass lower limit from \citet{mascareno_2021}) model which can also be ruled out at 30$\sigma$ confidence level. A hazy atmospheric model (tholin haze) with a haze formation efficiency of 0.01\% and 12~M$_{\oplus}$ can be ruled out at 38$\sigma$ confidence level as it produces a strong slope and a feature at 3.3$\mu$m which is not seen. The red dashed line shows a 12~M$_{\oplus}$, 10xsolar metallicity model with an internal temperature of 100~K, and can be ruled out at 18$\sigma$ confidence level as it produces a strong methane absorption feature at 3.3$\mu$m. The red solid line is the best fit model from the HST transmission spectrum \citep{barat2023}, extrapolated to JWST wavelengths. It has 23~M$_{\oplus}$, solar metallicity, T$_{int}$=400~K. This model is simulated using \texttt{PetitRadtrans}. This model matches with the water and methane features in NRS1, however underpredicts the size of the CO$_2$ feature, most likely due to a higher mass, lower metallicity and no photochemistry. All the models presented in this figure have a grey cloud deck at 0.01 bar. }
      \label{fig:spectra_full}
\end{figure*}

We also applied the same free retrieval framework on the transmission spectrum derived from the SPARTA reduction. We find the same molecular detections as found with the Eureka! reduction and the VMR are consistent within 1$\sigma$. The retrieved planet mass (13$\pm$1.5~M$_{\oplus}$) from SPARTA reduction is also consistent within 1$\sigma$.

\subsection{Forward Modeling and gridtrievals on the transmission spectrum of V1298 Tau b} \label{grid models}
In addition to using free PICASO retrievals (Section \ref{spectrum}), we model the observed spectra of V1298 Tau b using two independent grids of self-consistent forward models generated using PICASO and ATMO. While free chemistry retrieval has the flexibility to modify its chemical abundances to fit the data (i.e., it is data driven), the self-consistent models take into account the physical and chemical processes of the atmosphere (i.e, it is theoretically driven). 

\subsubsection{PICASO self-consistent grid} \label{picaso}

We compute a radiative--convective--thermochemical--equilibrium (RCTE) grid of forward models using the 1D climate model \texttt{PICASO} \citep{Mukherjee22}. The {\it Photochem} 1D chemical kinetics model \citep{wogan23} was then used to post-process chemical disequilibrium effects such as vertical mixing and photochemistry on this grid to forward models. The chemistry grid was sampled for metallicity, C / O ratio, internal temperature (T$_{\rm int}$) and eddy diffusion coefficient (K$_{\rm zz}$), resulting in $\sim$4000 models. Further details of the PICASO forward models are provided in Appendix \ref{appendix: forward models}.

We also varied the planet mass, gray cloud opacity, and reference radius while computing the transmission spectrum from the chemistry grids. We include a temperature offset while fitting the grid models, which is a constant additive term to the T-P profile to account for any deviation from the self-consistent T-P profile. The posterior distribution obtained from the \texttt{PICASO} grid fitting setup is shown in Figure \ref{fig:grid_corner}. Since the number of grid models is smaller compared to atmospheric retrievals and is limited by the resolution of the grids, we do not estimate uncertainties on the atmospheric parameters from the grid fits using Bayesian inference methods. Therefore, we assign probabilities to each grid model by comparing with the observations (e$^{-{\chi}^2/2}$) and calculate the weighted average and weighted standard deviation for each parameter. The values of the best-fit grid model and uncertainties are given in Table \ref{tab:retrieval table}. We note that the quoted uncertainties should be treated as a first-order approximation, and often these uncertainties would change when we change the grid resolution.

The best-fit model from the PICASO grid shows a super-solar atmospheric metallicity (logZ/$Z_{\odot}$=1.05$\pm$0.05) and sub-solar C/O ratio (0.23$\pm$0.08). The metallicity and C/O ratio obtained from the grid fit is consistent with the metallicity and C/O ratio calculated from the free PICASO retrieval within 1$\sigma$ (see Appendix \ref{grid models} and Figure \ref{fig:metal_co_grid_free_comparison}). Our models find a best-fit mass of 15$\pm$1.7~M$_{\oplus}$, which is consistent with the free PICASO retrieval at the confidence level of 2 $\sigma$.

\begin{figure}
    \centering
    \includegraphics[width=1\linewidth]{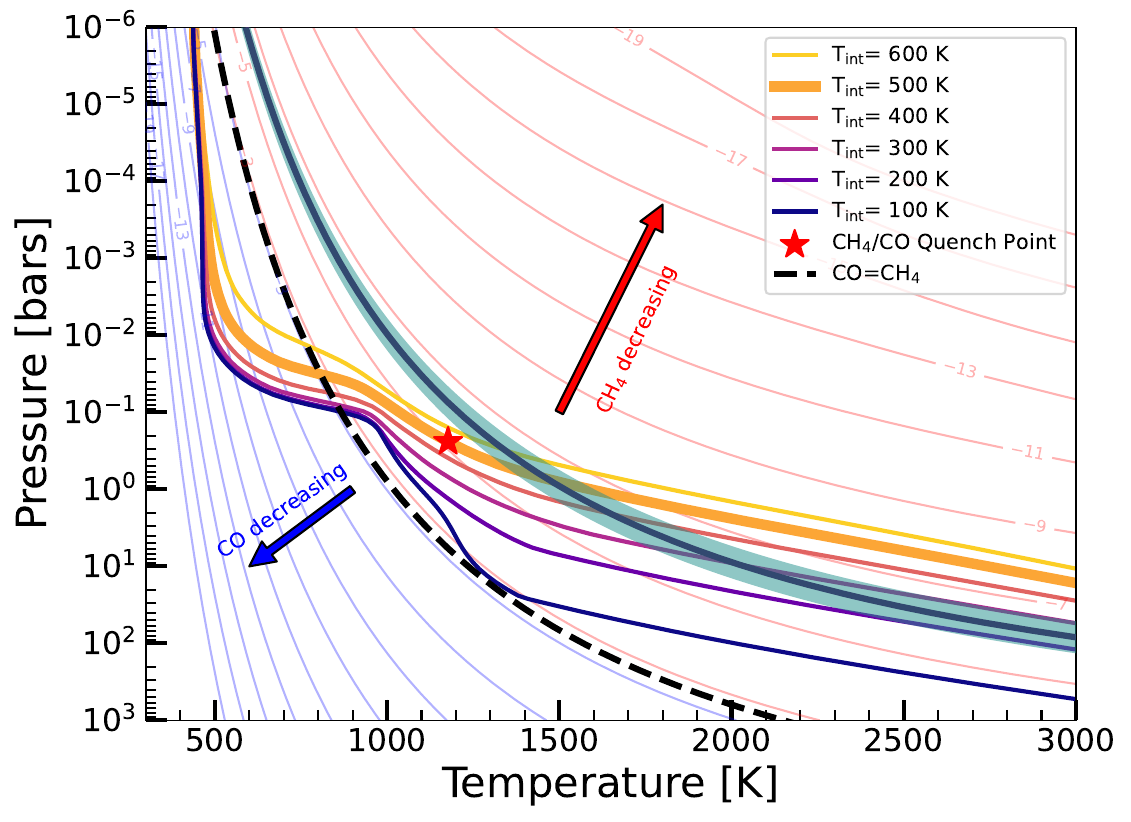}
    \caption{Plausible self-consistently calculated T-P profiles for V1298 Tau b for different values of T$_{int}$ using PICASO. The light blue and red lines correspond to constant abundance CO and CH$_4$ contours, respectively. The black dashed line shows the CO/CH4 equality line for 10$\times$solar metallicity \citep{fortney_2020}. The solid blue line with its shaded region  shows the methane abundance derived from the free retrieval and its 1$\sigma$ uncertainty. The red star shows the CH$_4$/CO quench point for the T-P profile with T$_{int}$=500~K. }
    \label{fig:co/ch4_line}
\end{figure}

To explain the apparent low abundance of methane, our forward models require the deep atmosphere to have a high temperature and vertical mixing. The PICASO grid finds a relatively high internal temperature ($\sim$500~K), compared to the expected internal temperature between 100-150~K for a Neptune or lower mass at this age. We also find a relatively low vertical mixing coefficient ($<$10$^7~cm^{2}/s$). From our forward models, we find that self-consistent T-P profiles with T$_{int}$ lower than 200~K do not cross the required methane abundance contour to explain our observations. For T-P profiles with T$_{int}$ between 200-400~K, the required quench point lies deep within the atmosphere (1-10 bar) and requires unrealistically high K$_{zz}$ values. For such high K$_{zz}$ values, the CO$_2$ abundance is also significantly affected and cannot fit our transmission spectrum. Therefore, the combination of methane and CO$_2$ constrain the upper bound on K$_{zz}$, and thus T$_{int}$.

Initially, we had computed the grid using full day-to-night heat redistribution, but in that case we found that the temperature offset was large ($\sim$-200K), making the isothermal part of the atmosphere consistent with the temperature recovered from the free chemistry retrievals. Since atmospheric chemistry is strongly dependent on temperature, we recomputed the grid using lower (\texttt{rfacv}$\sim$0.2) day-to-night heat redistribution, which implies a colder day-night terminator compared to full day-to-night heat redistribution. For the recomputed grid, the temperature offset is consistent with 0. The atmospheric metallicity, C/O ratio, mass, and internal temperature constraints are not affected significantly between the two grids discussed here. We present the results of the latter grid models here. The grid fit converges to $\sim$400$\pm$10~ppm offset between the HST and JWST spectra (consistent with the prediction from the known optical variability of the star); however, we do not find a significant offset between NRS1 and NRS2, which is consistent with free atmospheric retrievals. 

\subsubsection{ATMO self-consistent grid} \label{atmo_model_description}

We also computed a forward model grid of radiative--convective--thermochemical--equilibrium (RCTE) $P$-$T$ profiles using ATMO \citep{Tremblin2015, Drummond2016, Goyal2020} to compare with the results from the PICASO forward model grid (Section \ref{picaso}). 

The ATMO grid consists of a range of planet mass (5M$_{\oplus}$, 15M$_{\oplus}$, 30M$_{\oplus}$), recirculation factor (0.25, 0.5, 0.75, 1.0), metallicity (0.1$\times$, 1$\times$, 5$\times$, 10$\times$, 100$\times$ solar), C/O ratio (0.05, 0.1, 0.2, 0.35, 0.55, 0.70, 1.0) and internal temperature (200~K, 400~K, 600~K). For each of the RCTE T-P profiles, we calculate the chemical abundance profiles using VULCAN \citep{tsai2023}. We implement disequilibrium chemistry (vertical mixing and photochemistry) using VULCAN. We compute two sets of grids; one including only vertical mixing and another including both vertical mixing and photochemistry.

To compute the chemistry grid, we consider a range of K$_{zz}$ values (10$^6$, 10$^8$, 10$^{10}$). The observed stellar spectra of V1298 Tau \citep{Duvvuri2023} were used to initialize the photochemistry in the model. Using these chemistry grid, we generate a grid of transmission spectra for V1298 Tau b. We include a power-law continuum opacity function to include the effect of clouds/hazes. Spectra were generated for a range of haze factors (1$\times$ and 150$\times$ nominal Rayleigh scattering) and gray cloud opacity (0$\times$, 0.5$\times$, 1$\times$ H$_2$ Rayleigh scattering cross section at 350~nm) as described in \citet{Goyal2018}. The final grid of transmission spectrum with RCTE $P$-$T$ profiles and dis-equilibrium chemical abundances has more than 11000 model spectra.  Further details of the ATMO forward model grid are provided in Appendix \ref{appendix:atmo_model_descriptions}.

Using this grid of transmission spectra, we fit the combined HST and JWST transmission spectrum of V1298 Tau b, using the techniques of ${\chi}^{2}$ minimization. We include a constant offset between HST and JWST spectra as an additional parameter while fitting, similar to the PICASO models. We find the best fit planet mass of 15M$_{\oplus}$, recirculation factor of 0.5, 10$\times$ solar metallicity, C/O ratio of 0.35, internal temperature of 600 K,  K$_{\rm zz}$ of 10$^8$, haze factor of 1$\times$ the nominal Rayleigh scattering and cloud factor of 1$\times$ from both the grids (vertical mixing only and vertical mixing with photochemistry). The mass, metallicity, C/O ratio and T$_{int}$ grid spacing is relatively coarse compared to the PICASO grid. Therefore, we do not calculate uncertainties from the ATMO grid as the uncertainties are dominated by the grid spacing and are not a true representative of the real parameter uncertainty.

\begin{figure}
    \centering
    \includegraphics[width=1\linewidth]{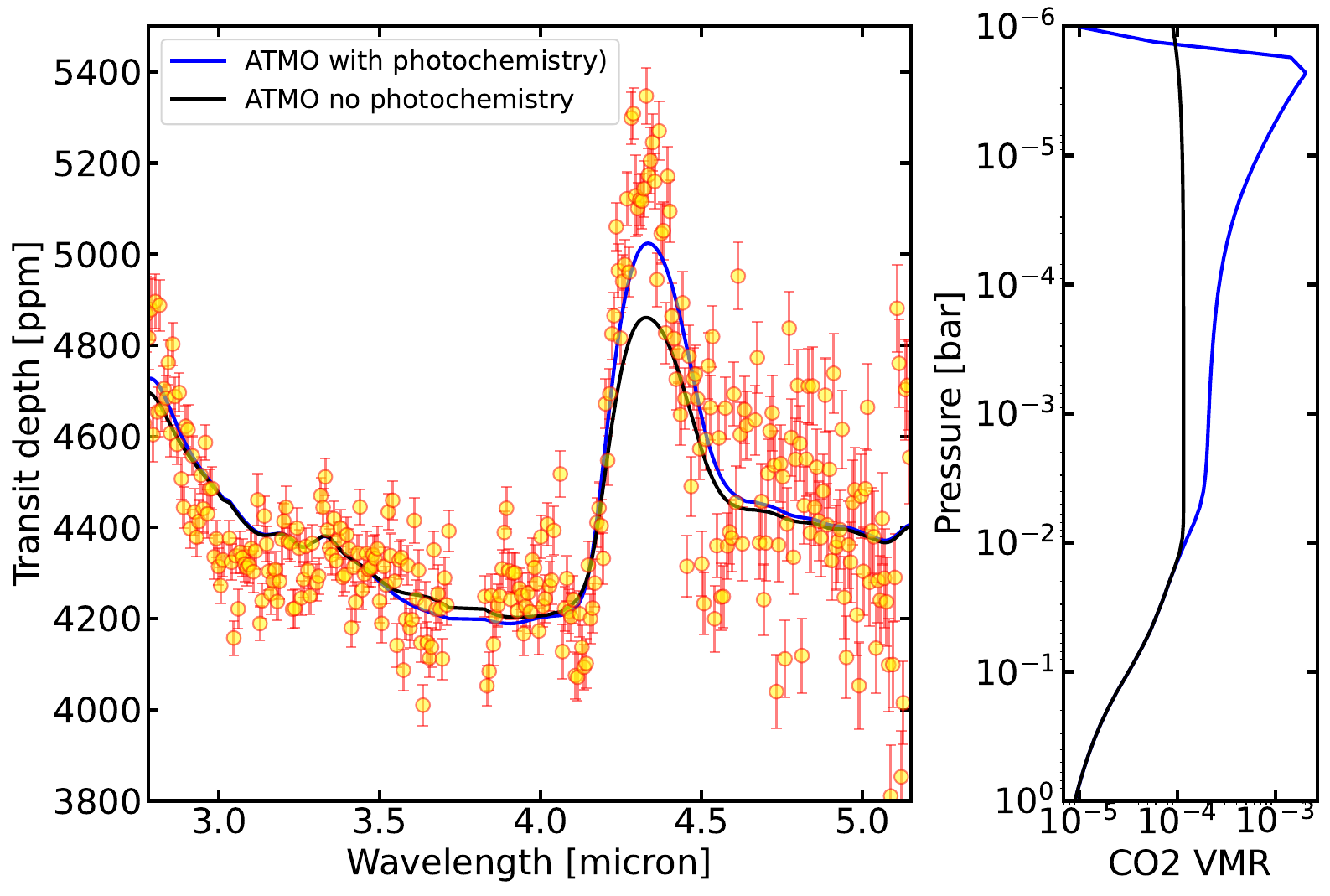}
    \caption{The observed JWST/NIRSpec transmission spectrum of V1298 Tau b (yellow points with $1\sigma$ errorbars) overplotted with best fit models from ATMO grids with photochemistry (blue continuous curve) and without photochemistry (black continuous curve), as described in Section \ref{atmo_model_description}. Vertical mixing is included in both grids. The right panel shows their corresponding CO$_2$ VMR profiles. Both models have the same metallicity, C/O ratio, K$_{zz}$ and planet mass. The model including photochemistry provides a better fit of the CO$_2$ feature. The photochemistry model does not affect the size of the absorption features for other molecules significantly.   }
    \label{fig:atmo_models}
\end{figure}

In Figure \ref{fig:atmo_models} we show the best-fit models from both the ATMO grids. We find that including photochemistry increases the CO$_{2}$ abundance by $\sim$0.3 dex at 10$^{-4}$~bar, resulting in an overall increase in the CO$_{2}$ feature by $\sim$100~ppm and fits the observed spectra better. The $\chi^2$ value for the fit with photochemistry and vertical mixing is 2081 compared to 2444 for the fit with just vertical mixing, for 359 observational data-points (bins). The other molecular features are similar between the two grids. This is consistent with previous studies that have shown the effect of photochemistry on the CO$_{2}$ abundance, especially for planets with equilibrium temperature lower than 600~K \citep{baeyens2022}.

In Table \ref{tab:retrieval table} we compare the retrieved parameters from ATMO grid with PICASO grid. The atmospheric parameters retrieved (T$_{int}$, K$_{zz}$ metallicity, cloud opacity) and the offset between HST and JWST data points are consistent within 1$\sigma$ between PICASO and ATMO grid fits. However, the retrieved C/O ratio from the ATMO grid is higher than the PICASO grid. We also note that the best-fit grid model with photochemistry from ATMO underestimates the CO$_2$ feature size by $\sim$100~ppm, which is not the case for the PICASO grid. 

By comparing the PICASO and ATMO grids we identify two key differences: they use different photochemistry models which could differ in terms of their chemical network, and the two grids use different algorithms to change C/O ratio. In PICASO the C/O ratio is decreased by decreasing the [C] and increasing the [O] keeping the [C+O] conserved \citep{Mukherjee22}. On the other hand ATMO only changes [O] to change the C/O ratio. Depending on how C/O ratio is changed in a forward model, for the same C/O ratio there could be significant changes to the abundances of all C and O bearing species such as H$_2$O, CO$_2$, CO and CH$_4$ and therefore affect the final transmission spectrum and its interpretation \citep{verma2025}. A detailed investigation is needed to understand how the method of changing C/O ratio affects the final transmission specturm and is out of the scope of this paper.

\begin{table*}
    \centering
      \resizebox{\textwidth}{!}{ \begin{tabular}{c|c|c|c}

    \hline
    \hline
        Parameter & PICASO Free retrieval (Section \ref{spectrum}) & PICASO 1D self-consistent grid (Section \ref{grid models})  & ATMO 1D self-consistent grid (Section \ref{atmo_model_description}) \\

        \hline
       T$_{int}$ [Kelvin]  & -- & $>$ 500$\pm$50 &   600 \\  
     
      log Z [Z$_{\odot}$]   & 0.6$^{+0.4}_{-0.6}$ & 1.05$\pm$0.2  & 1.0 \\
      
       C/O   & 0.22$^{+0.06}_{-0.05}$ & 0.23$\pm$0.08  &   0.35 \\
       
       logK$_{zz}$ [cm$^2$/s]  & -- & 7$\pm$0.9 &   $<$10$^8$ \\
       
       Mass [M$_{\oplus}$]  & 12$\pm$1 & 15$\pm$1.7  &   15 \\
       
       log${\kappa}_{cld}$ [cm$^2$/g]  & -2.77$\pm$0.23 & -2.67$\pm$0.22 & -2.80  \\
       
       HST offset [ppm] & 405$\pm$15 & 397$\pm$10 &   400\\

       \hline
    \end{tabular}}
    \caption{Table showing retrieved atmospheric properties from PICASO (free and self consistent grid) and ATMO (self-consistent forward grid). We note that the values shown for the ATMO grid are for both the vertical mixing only and the vertical mixing+photochemistry grids which converge to the same best-fit values. For details of the models see Section \ref{spectrum} and \ref{grid models}. The quoted grid values are taken from the best fit (least chi-square) grid model to the observed transmission spectrum. The uncertainties from PICASO grid models are estimated by assigning probabilities to each grid model and calculating weighted standard deviation. For the ATMO grid we do not compute uncertainties as the mass-metallicity-C/O ratio grid spacing is larger than PICASO and the uncertainty estimated from this coarse grid is not representative of the true parameter uncertainties. }
    \label{tab:retrieval table}
\end{table*}

\subsection{Modeling haze formation in V1298 Tau b} \label{haze formation}

V1298 Tau b receives a high amount of UV radiation from its young host star: $\sim$100$\times$solar \citep{Duvvuri2023}. UV radiation is known to be an important ingredient in photochemical haze formation \citep{kawashima2018,gao2020}  The equilibrium temperature is expected to be $\sim$ 600K. This temperature is close to the CH$_4$/CO transition temperature \citep{fortney_2020} and has been found to be right at the sweet spot for haze formation in the sub-Neptunes \citep{yu2021,Brande2024}.

We simulated transmission spectra of V1298 Tau b including atmospheric hazes using photochemistry, haze microphysics, and transmission spectrum models of \cite{kawashima2018} following the prescription outlined in \cite{kawashima2019a} and \citet{kawashima2019}. We assume a composition of Titan-like tholin haze \citep{khare1984}. We consider CH$_4$, HCN and C$_2$H$_2$ as haze precursor molecules in our model. We derive their abundances from photochemical models. We adopted the stellar spectra from \citet{Duvvuri2023}. The haze formation efficiency, i.e. the fraction of haze precursors which can form haze is an uncertain parameter. We simulate models with haze formation efficiencies between 0.01\%-1\%.  Further details of the haze models are provided in Appendix \ref{appendix: haze details}. 

We compare a model with 0.01\% haze formation efficiency and a mass of 12~M$_{\oplus}$ in Figure \ref{fig:spectra_full} with the observed spectrum of V1298 Tau b. The large absorption features and the absence of an apparent slope in the spectrum rule out the high abundance of photochemical hazes .

Haze formation has been found to be ubiquitous for mature sub-Neptunes \citep{desert2011a,kempton2012,morley2013,kreidberg14,knutson2014,libby_robert2020,kempton2023,bean2021}. Thus, the lack of atmospheric hazes for a young sub-Neptune progenitor hints that the atmospheric composition of sub-Neptunes undergoes significant changes as they evolve. Haze formation depends on atmospheric metallicity; high metallicity favors haze formation \citep{kawashima2019,kawashima2018}. Furthermore, methane has been considered as a precursor molecule for haze formation. The low abundance of methane and relatively low atmospheric metallicity of V1298 Tau b compared to typical sub-Neptunes could explain the low haze production on this planet.

\subsection{Interior structure evolution models of V1298 Tau b} \label{structure and evolution}

We generate a grid of interior structure and evolution models for V1298 Tau b following the formalism outlined in \citet{vazan2022}. The evolutionary model starts at the time of disk dispersal. The initial conditions are derived from the results of planet formation models \citep{ormel2021}.
Our models account for core-erosion and mixing with the envelope when applicable. We run these evolution models for two planet masses (10 and 15~M$_{\oplus}$), which are selected as representative values of the mass posterior distribution. We pick two values of atmospheric metallicity (1 and 10 times solar) given the constraints we derive from the atmospheric retrievals. The metallicity sets the envelope opacity and plays a key role in the cooling of the planet.

\begin{figure*}
    \centering
    \includegraphics[width=1\linewidth]{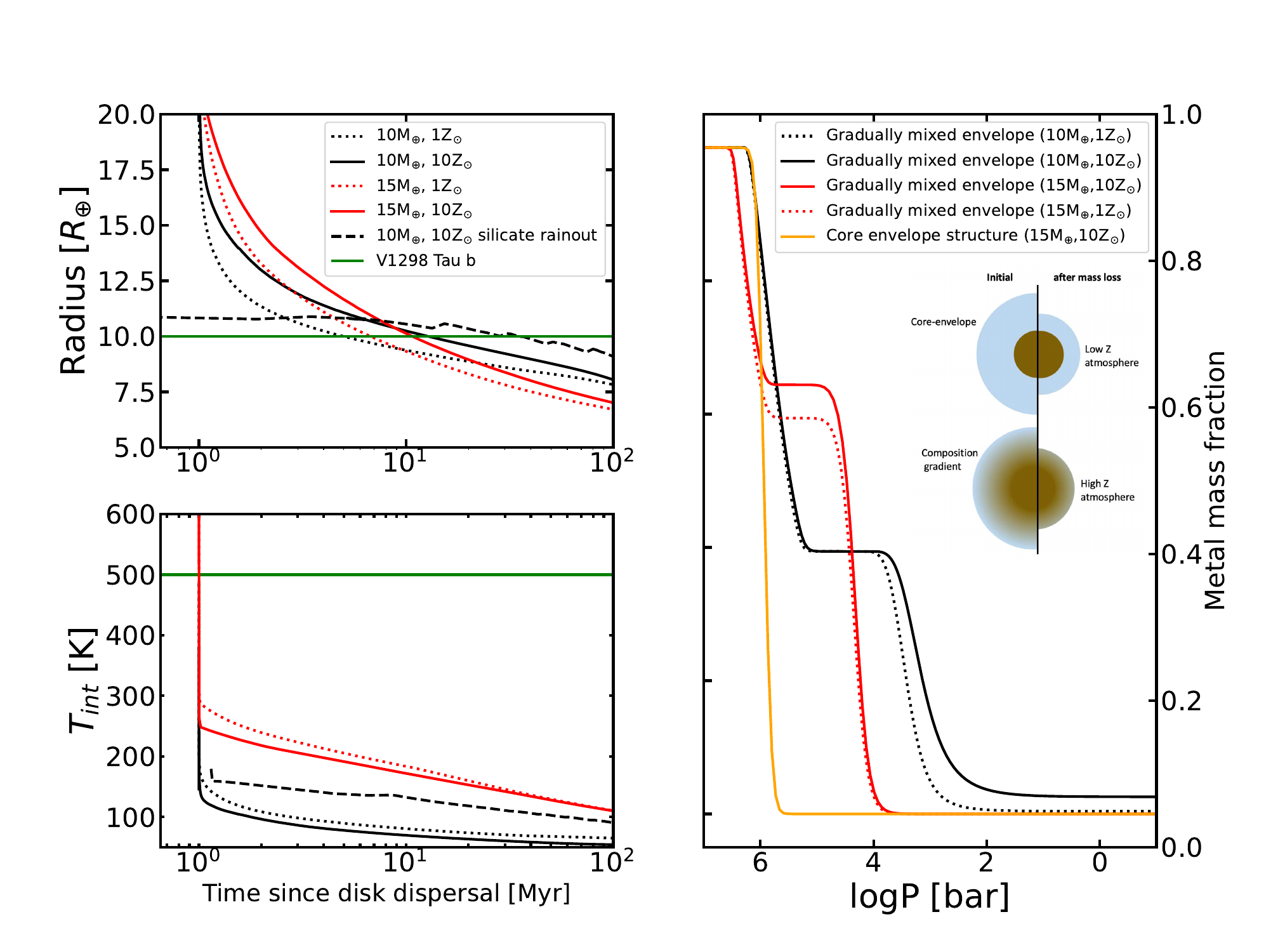}
    \caption{Evolutionnary track models of the radius  (upper left panel)and of the internal temperature (lower left panel) for V1298 Tau b considering various ranges of mass and metallicity. Models with two masses (10,15~M$_{\oplus}$) are calculated, and for each mass we show two models with solar and 10 times solar atmospheric metallicity. Black lines represent 10~M$_{\oplus}$ and red lines represent 15~M$_{\oplus}$. Dotted lines correspond to solar metallicity and continuous lines correspond to 10 times solar metallicity.  These ranges of mass and atmospheric metallicity values are chosen as representative and consistent within 3$\sigma$ of the posterior distribution from the atmospheric retrievals (Table \ref{tab:retrieval table}). The black dashed models shows radius and internal temperature evolution for a 10~M$_{\oplus}$, 10 times solar metallicity model including silicate rainout. The evolution tracks for V1298 Tau b have been simulated using interior structure and evolutionary models from \citet{vazan2018,vazan2024}. See Section \ref{structure and evolution} for further details on the models. The green continuous line shows the lower limit of internal temperature inferred from the atmospheric retrievals estimated from this work and the radius of planet b. The internal temperature we measure ($\sim$500~K) appears inconsistent with these evolution models, and requires alternative formalisms. These models require 20-50\% by mass H/He envelope, they have been simulated using a gradually mixed interior structure assuming a compositional gradient. In the right panel we show the atmospheric metallicity profiles that have been used for our evolutionary models presented in the left panel. The same linestyle and color scheme have been used. For a comparison, we also show a core-envelope structure   (orange continuous) line for a 15~M$_{\oplus}$, 10 times solar atmospheric metallicity model. At the current age, both core-envelope and gradually mixed envelope models can explain the observed composition of the upper atmosphere. Two globally different scenario could be considered:  the gradually mixed interior scenario, in which the atmospheric metallicity is expected to increase with age due to removal of upper layers of the envelope, whereas for the core-envelope structure scenario, the metallicity of V1298 Tau b is not expected to change over time, regardless of mass loss rates (See right panel inset).}

    \label{fig:interior_structure_model}
\end{figure*}

We calculated models with two different interior structures (gradually mixed interiors and core-envelope structure). In the core-envelope scenario, the core and the envelope are differentiated and, in the gradually mixed case, the metals are mixed into the atmosphere, resulting in a metallicity gradient in the envelope. There has been growing evidence of gradually mixed interiors or `fuzzy-cores' in our solar system giant planets \citep{wahl2017,2mankovich2021,vazan2020}. The different interior structure models are shown in Figure \ref{fig:interior_structure_model}, right panel. At the current age, the upper observable atmosphere looks similar for all interior structure models. We include atmospheric mass loss according to the energy-limited atmospheric escape approach, in which mass loss rate evolves with planetary radius, mass and distance from the star. The time dependent mass loss efficiency ($\eta$) and high energy luminosity (L$_{XUV}$) are adopted from \citet{rogersandowen2021}.

 We show the radius and internal temperature evolution in Figure \ref{fig:interior_structure_model} for two gradually mixed interior structures.   We derive the internal temperature by converting the luminosity of the planet which is an output from our evolution model at each time step of the simulation, assuming blackbody law. 

For the models shown in Figure \ref{fig:interior_structure_model}, the radius of the planet is best matched by 20-50\% by mass H/He envelope. These models show that the planet's age (i.e., the time since disk dispersal) could be between 5-15~Myr. The age shows a dependence on assumed atmospheric metallicity, and higher metallicity models show a higher age of the planet because higher opacity increases the cooling timescale. The inferred planet age is not strongly dependent on the planet mass. It is important to note that for different assumptions, such as lower atmospheric mass fraction or initial entropy, the inferred age of the planet could be closer to 20~Myr (See Figure \ref{fig:evolution_comparison}). 

If V1298 Tau b has an envelope mass at the high end of the possible range, it is very close to the run-away gas accretion regime \citep{pollack1996,boddenheimer1986}, and must have stopped accretion at the cusp of entering this run-away phase. On the other hand, we find the internal temperature for this planet to be high ($\sim$500K) from our self-consistent atmospheric grids (Section \ref{grid models}), while evolutionary models show that even if the planet is only 10~Myr old with a mass of 15~M$_{\oplus}$ (3$\sigma$ upper limit from atmospheric retrievals), its internal temperature is expected to be $\sim$200~K. The rainout (condensation and settling) of silicates in the atmospheres of low-mass planets has been theorized to inflate the envelope and release energy of gravitational settling and condensation \citep{vazan2024}. We show a model for V1298 Tau b, including silicate rainout in Figure \ref{fig:interior_structure_model}. This model produces an internal temperature almost twice that of those without rainout; however, it is still not sufficient to match the extremely high internal temperature we infer for this planet. 

Therefore, internal temperature constraints from the atmospheric models are inconsistent with theoretical evolutionary models. As the cooling timescale for such a high internal temperature is of the order of 1000 years, explaining the internal temperature using energy of formation, would require the planet to emerge from the disk in the last few thousand years. This is unlikely given that the timescale for inner disk dispersal is on the order of 10$^{5}$ yr \citep{clarke2001}, which is much longer than the necessary cooling timescale. Even if we consider that the star is $\sim$10~Myr old \citep{finoceti2023,maggio2022}, such a short cooling timescale is inconsistent with the age of the system.

\section{Discussion} \label{section:discussion}

\subsection{Limb darkening of V1298 Tau}

\begin{figure}
    \centering
    \includegraphics[width=\linewidth]{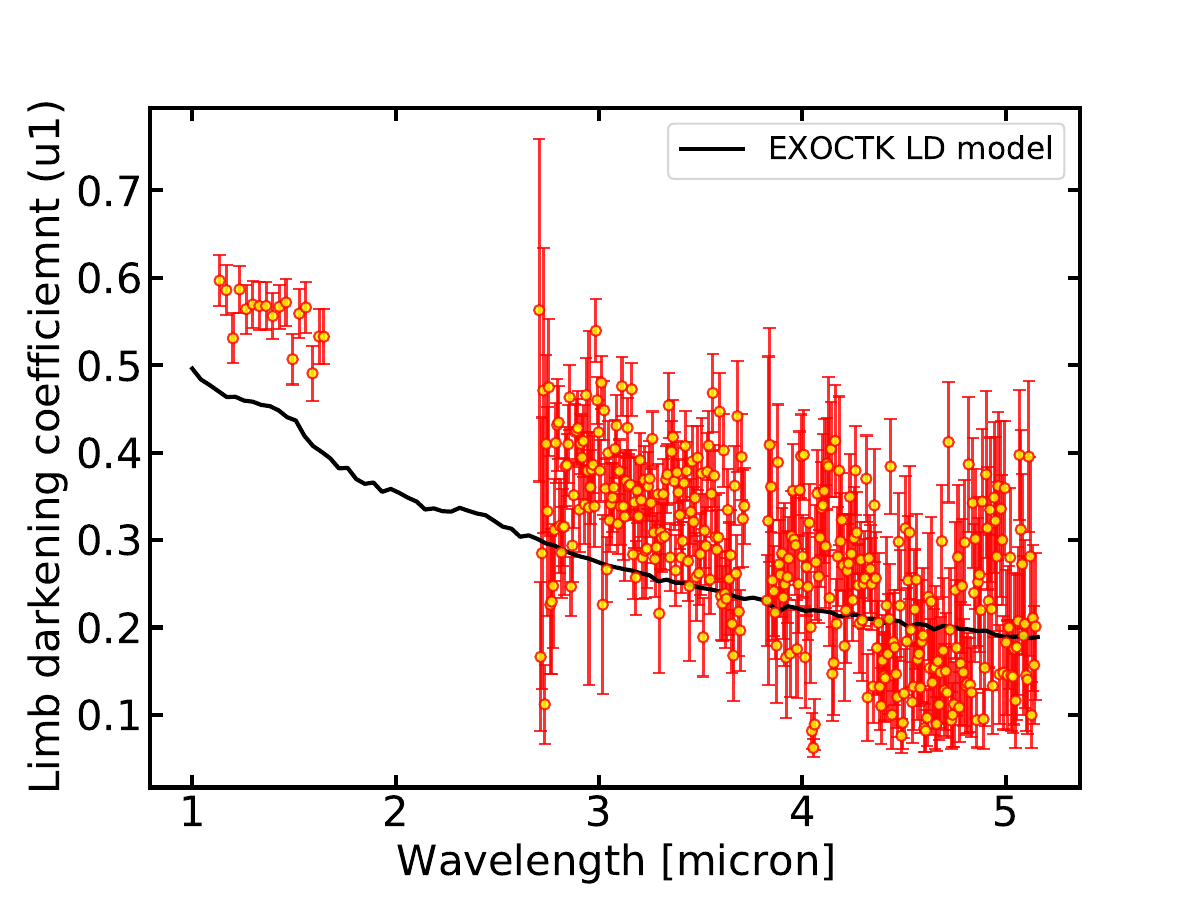}
    \caption{Fitted linear limb darkening coefficient for V1298 Tau from the HST and JWST transit observations of V1298 Tau b (yellow data points with red $1\sigma$ errobars). The JWST limb darkening coefficients are measured at a higher spectral resolution compared to HST (by a factor of $\sim$10), which is why the JWST errorbars are larger compared to HST. The continuous black curve is from a linear limb darkening model generated using EXOTIC-LD \citep{exotic_ld} computed for a K1 type main sequence star. The modelled linear limb darkening coefficient are found to be significantly smaller than the observed values, but the difference gets smaller at longer wavelengths.}
    \label{fig:limb_darkening}
\end{figure}

Limb darkening models, such as Exotic-LD \citep{exotic_ld} can calculate limb darkening for main-sequence stars based on stellar structure models. However, these models may not be applicable for young T-Tauri stars. Limb darkening is known to be affected by the magnetic field of the star \citep{kostogryz2024}, and V1298 Tau is known to have a magnetic field strength between 100-300 Gauss \citep{finoceti2023}. HST observations of V1298 Tau have shown greater limb darkening compared to main sequence models \citep{barat2024b}. 

In Figure \ref{fig:limb_darkening} we show the limb darkening coefficients from the HST and JWST light curve fits. We find that the offset between the model and the fitted limb darkening coefficients decreases at longer wavelengths. However, there seems to be a dip in the fitted coefficients around 4.4$\mu$m. 

We investigate the effect the feature around 4.4$\mu$m for the limb darkening has on the transmission spectrum. We note that the limb darkening and the flare amplitude are correlated. So we run the spectroscopic light curve fits for four cases: i. limb darkening coefficient fixed to EXOCTK model and flare amplitude left free ii. limb darkening free but flare amplitude fixed to flare amplitude derived from empirical model fit and smoothed with a gaussian kernel iii. both fixed iv. both free. However, we find that the spectra derived from all these cases are qualitatively similar and agree with each other at the 1$\sigma$ level for all spectroscopic channels. We conclude that the feature in the limb darkening function does not affect the transmission spectrum significantly.

We see that the difference between the model limb darkening and the fitted is higher for bluer wavelengths which could be due to inhomogeneities on the surface of the star. If this is a spot induced effect the difference should be decreasing with longer wavelengths as the contrast between the photosphere and spots decrease at these wavelengths.

\subsection{Comparing and combining the HST and JWST transmission spectra of V1298 Tau b}

From the HST/WFC3 G141 transmission spectrum \citet{barat2023} reported a large scale height from a water absorption feature and a non-detection of methane. The HST spectrum can be explained using a solar metallicity atmosphere with a Neptune-like mass. In Figure \ref{fig:spectra_full} we show the best fit model for HST spectrum of V1298 Tau b from \citet{barat2023}, extrapolated to JWST wavelengths. The extrapolated model matches the water and methane features in the JWST band. It however underpredicts the CO$_2$ feature. The HST models were for a higher planet mass (23~M$_{\oplus}$) and lower metallicity (solar) compared to the constraints we derive from JWST. CO$_2$ is strongly affected by atmospheric metallicity. Furthermore, the CO$_2$ is affected by photochemistry (Figure \ref{fig:atmo_models}) which was not included in these models.

From our free chemistry and grid fits (PICASO and ATMO) we find an offset of $\sim$400$\pm$10 ppm. The expected offset derived from the optical photometry observations is 390 ppm assuming an $\alpha$=-1.75. Therefore, the retrieved offset between JWST and HST is comparable to the expected value from the observed optical variability. It shows that given the large scale height of the planet, stellar activity induced offsets between different epochs can be retrieved from the observations itself.

Given the differences in the HST model (lower metallicity, higher mass, different K$_{zz}$, and no photochemistry) the transmission spectrum predicted for JWST from these models and adjusted for stellar activity induced offsets is consistent with our observed JWST transmission spectrum. Therefore, we can combine the HST and JWST transmission spectra, which we have done for our analysis to determine the mass and atmospheric composition of V1298 Tau b.

\subsection{Comparison between free retrievals and the PICASO grid}

\begin{figure}
    \centering
    \includegraphics[width=\linewidth]{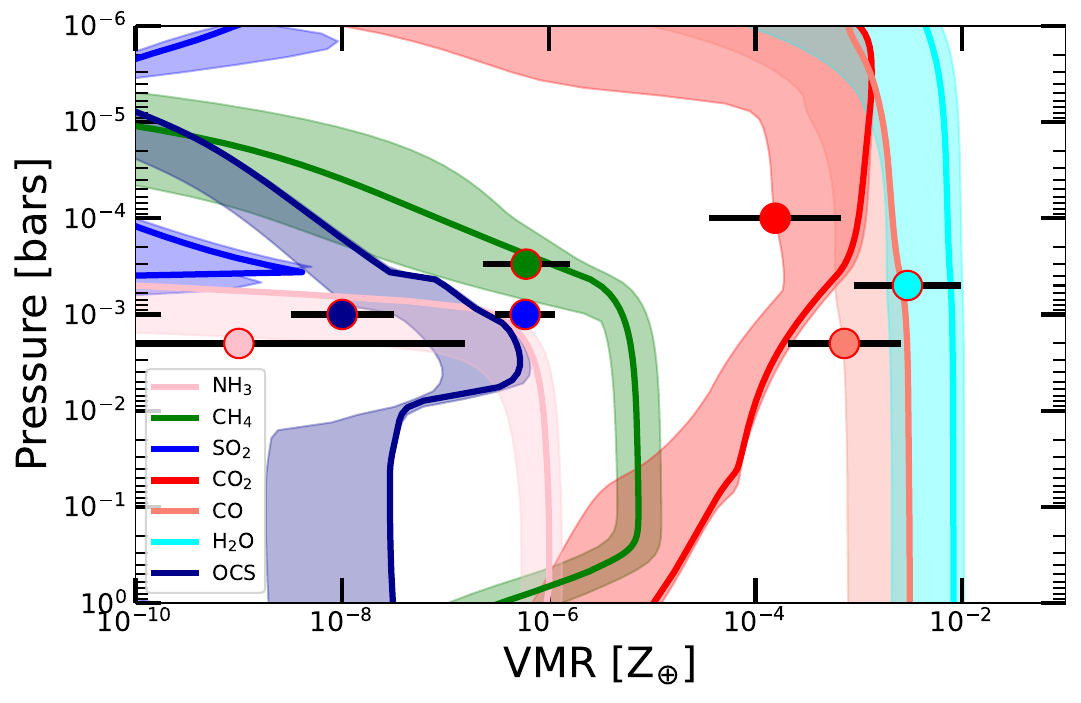}
    \caption{Comparison between retrieved abundances of molecules in our free chemistry  PICASO retrieval (Appendix \ref{appendix: free retrieval}), shown with colored circles and 1$\sigma$ uncertainties, with molecular abundance profiles (solid lines of same color) from best fit self-consistent PICASO grid model (Appendix \ref{appendix: forward models}). The shaded region around each profile from our grid represents the range of abundance profiles in the grid models which have a probability $p(x)\ge$1\% (3$\sigma$ confidence region). In our case, this represents the top 61 best-fitting grid models. Although the free retrieval finds SO$_2$, the best fit grid model does not seem to have a large SO$_2$ abundance. The grid predicts OCS, which is also found in the free retrievals but at a 3$\sigma$ lower VMR than predicted by the grid.}
    \label{fig:comparison_free_grid}
\end{figure}

We compare the derived molecular abundances from the free retrieval and the PICASO grid fit in Figure \ref{fig:comparison_free_grid}. 
The abundance profile of the best-fitting PICASO grid model is shown as solid lines, whereas the shaded region around each profile represents the range of abundance profiles in the grid models that have a probability $p(x)\ge$0.001\%. In our case, this represents the top 104 best-fitting grid models.  H$_2$O, CH$_4$, CO, and CO$_2$, OCS are consistent between the best-fit PICASO grid model and the retrieved values from the PICASO free chemistry retrieval. The metallicity and C/O ratio we derive from the free chemistry ([M/H]=0.6$^{+0.4}_{-0.6}$ and C/O ratio=0.22$^{+0.06}_{-0.05}$ respectively) are consistent within 1$\sigma$ with the best fit values from the PICASO grid (1.0$\pm$0.2 and 0.23$\pm$0.08 respectively). A comparison of the metallicity and C/O ratio from the free retrieval and the PICASO grid are shown in Figure \ref{fig:comparison_free_grid}. The mass derived from the free retrieval (12$\pm$1~M$_{\oplus}$) is also consistent within 2$\sigma$ with the best fit mass of the PICASO grid. 

The free retrieval can measure the molecular abundances independently (without any physical constraints), and therefore estimates the metallicity and C/O ratio completely driven by the data. However, the self-consistent grid is not as flexible as the free retrieval but takes into account physical and chemical processes within the atmosphere. The agreement between these two independent and complementary approaches shows the robustness of the inferred atmospheric composition.

Our free retrieval finds SO$_2$ at 4$\sigma$ level confidence, but the self-consistent PICASO grid does not predict SO$_2$. Since the isothermal temperature of the self-consistent grid is low ($\sim$450~K) its predicted SO$_2$ abundance is low, as SO$_2$ decreases for lower temperature \citep{mukherjee2024}. To test the robustness of the SO$_2$ detection, we extract a 20 pixel bin transmission spectrum for V1298 Tau b and re-run our free retrieval. In this case, we do not find SO$_2$. It could be that SO$_2$ is narrow and larger spectral bins could result in dilution of the feature. However, for the 20-pixel-bin transmission spectrum, we retrieve the OCS from free retrieval at a similar VMR ($\sim$10$^{-8}$ VMR) compared to the 10-pixel-bin case. \citet{mukherjee2024} predict the onset of OCS at temperatures lower than 500~K. Thus, this test shows that the OCS detection from the free retrieval, although weak (3.5$\sigma$), is consistent with predictions from the PICASO grid.

The equilibrium temperature from the free chemistry retrieval ($\sim$450K) is lower than the expected equilibrium temperature of the planet ($\sim$670K). We found that grids with full heat redistribution required $\sim$ 200~K temperature offset to explain the observations. Therefore, in our final models we present results from a PICASO grid with a lower heat redistribution such that it matches the temperature obtained from the free chemistry retrievals. Our models prefer a lower day-night terminator temperature compared to the equilibrium temperature to explain the observed transmission spectrum.

\subsection{Determining the mass of V1298 Tau b from the transmission spectrum} \label{section: planet mass}

The mass of an exoplanet can be constrained by its transmission spectrum due to the dependence of the scale height on surface gravity \citep{dewitt_seager2013, batalha19}. It has been shown that including mass as free parameter in atmospheric retrievals still allows us to measure atmospheric molecular abundances accurately for primordial H/He atmospheres using broadband transmission spectrum \citep{maio2023}.

Therefore, we include mass as a free parameter in both free chemistry retrieval and self-consistent grid fits with PICASO and ATMO performed on the combined HST and JWST spectra of V1298 Tau b. Our models measure a mass of 12$\pm$1~M$_\oplus$ from the free retrieval. The grid fits with PICASO and ATMO independently constrains the mass to 15$\pm$1.7~M$_\oplus$ and 15~M$_{\oplus}$ respectively.  We note that the uncertainties calculated on the mass from the PICASO grid are not derived from a Bayesian retrieval framework,  because the grids are coarse, but rather by assigning probabilities ($e^{-{\chi}^2/2}$) to each grid model. The masses derived from the two grids and the free retrieval are shown in Table \ref{tab:retrieval table}. 

The mass constrained from the combined JWST and HST transmission spectrum in this work is in agreement with previous estimates from the HST spectrum alone \citep{barat2023,barat2024b} and rules out the best fit RV mass from \citet{mascareno_2021} at $\sim$40$\sigma$ level of confidence (Figure \ref{fig:spectra_full}). This mass measurement confirms V1298 Tau b as a sub-Neptune mass planet and combined with its large scale height implies a mean molecular weight consistent with a H/He composition. Therefore, V1298 Tau b is consistent with the gas-dwarf formation scenario scenario proposed for sub-Neptunes \citep{rogers2025}.  Ongoing TTV studies of this system (Livingston et. al. in prep) also find a mass which is consistent with the mass estimates from the free retrieval and the grids within 1$\sigma$.

 Since the planet has a large scale height ($\sim$1500~km), the gravity is expected to vary significantly at the pressures probed by the water and C$O_{2}$ features. Therefore, we include non-constant gravity in our PICASO free retrieval, as well as both the PICASO and ATMO self-consistent forward model grids. We fit both the radius and gravity at a reference pressure. We also calculate an estimate of the uncertainty due to the assumption of a plane-parallel atmosphere (i.e neglecting any 3D effects) given the very large scale height. We find that uncertainty in estimating the scale height and therefore the planet mass from the 4.3$~\mu$m CO$_2$ feature, 1.4$~\mu$m and 2.7$~\mu$m water features independently is 5, 2, 3\% respectively. These estimated uncertainties are lower than the reported uncertainty from the free retrieval.

To further test the robustness of our method to retrieve the planet mass from the atmospheric scale height, we apply the same method on the NIRSPec/G395H spectrum of WASP-107b \citep{sing2024}. We chose this planet because its spectrum looks remarkably similar to V1298 Tau b. From a free chemistry retrieval we find a mass of 32$\pm$3~M$_{\oplus}$. Since WASP-107b is a quiet Sun-like star, there are reliable RV measurements. The RV mass for WASP-107b is 30.5$\pm$1.7~M$_{\oplus}$ \citep{piaulet2021}. Thus the mass we retrieve from the transmission spectrum of WASP-107b is consistent within 1$\sigma$ to the RV mass estimate. We also note that the uncertainty retrieved on the mass for WASP-107b with our proposed method is $\sim$10\%, and comparable to the uncertainty retrieved for V1298 Tau b ($\sim$8\%).

Transmission spectroscopy is a relative measurement, so changes to the absolute transit depth due to the transit light source effect \citep{rackham2019} is not expected to change the mass measurement significantly. This is especially true for hotter stars, such as K stars, where even for extreme cases the cold spot temperatures are higher than 3000~K.  This means that the contamination due to unocculted spots are not expected to alter molecular features such as water and CO$_2$ for stars like V1298 Tau, unlike the case for M-dwarfs \citep{barclay2021}.

  Measuring the mass from transmission spectra, while challenging and not advantageous compared to dynamical studies for most of the known exoplanets, is competitive when applied to variable T-Tauri stars which exhibit significant jitter in their RV signals. TTVs are an alternative; however, they also require long observing baselines and the systems to be near MMR. Thus, transmission spectroscopy could be a relatively inexpensive and efficient solution for mass characterization for young transiting planets.

\subsection{The methane thermometer}

\citet{mukherjee2024} have shown that the abundance of methane in the atmosphere of warm planets is dependent on the metallicity, C/O ratio, K$_{zz}$ and T$_{int}$. However, metallicity, C/O ratio and K$_{zz}$ affect the abundance of other molecules like H$_2$O, CO, CO$_2$ and SO$_{2}$ which are well constrained for V1298 Tau b. Thus, to explain the low methane abundance and also not affect the abundance of the other molecules our models require high temperature in the deep atmospheres. 
Therefore, our observations we are essentially sensitive to the temperature of the atmosphere near the methane quench point. 

However, to infer the planet's internal luminosity from the temperature of the deep atmospheric layers, we must assume an adiabatic T-P profile, implying a convective interior. If we assume the interior of the planet to be well mixed we can assume that large scale convection could be happening in the deep atmosphere. However, this assumption has been questioned for solar system giant planets \citep{wahl2017,2mankovich2021,vazan2022}, where observations suggest metallicity gradients. This idea is further discussed in detail in Section \ref{double diffusion convection}. In this case, even if we have a hot deep atmosphere, does not necessarily imply that the planet's luminosity is high.

\subsection{Interpreting the atmospheric composition of V1298 Tau b in the context of planet formation models} \label{low_co_insitu}

The transmission spectrum of V1298 Tau b reveals a primordial atmosphere rich in H / He, but with relatively low atmospheric metallicity ($\sim$10$\times$solar) and sub-solar C/O ratio ($\sim$0.2). Recent planet population synthesis models of gas giants by \citet{penzlin2024} show that low C/O ratios such as 0.2 are only found in aligned hot Jupiters which are believed to have migrated through the disk and accreted from the innermost parts of the disk ($\sim$0.2~AU). The innermost disk solids are carbon-depleted and are mostly made of oxygen-rich silicates that could lower the overall C/O ratio of the atmosphere, assuming that the silicates release their oxygen in the atmosphere. 

Extrapolating the same idea to V1298 Tau b, we could qualitatively say that this planet is likely to have accreted significantly from the inner part of the disk, which supports the in situ formation of the sub-Neptune/super-Earth model \citep{lee2014,lee_chiang2016}. The relatively low atmospheric metallicity is also consistent with formation within the water ice line. Otherwise, the planet could accrete a significant amount of volatile rich solids from beyond the ice line and could even become water worlds \citep{bitsch2021}.

\subsection{Effect of high internal temperature on the inferred atmospheric mass fraction and early evolution} \label{early evolution}

\subsubsection{Scenario 1: pure H/He atmosphere} \label{pure H atmo}

The interior structure models described in Section \ref{structure and evolution} show that the mass and radius of V1298 Tau b can be explained by the atmospheric mass fraction between 20-50\%, depending on the mass of the planet and the atmospheric opacity assumed.  However, these models predict a much lower internal temperature than inferred from our observations (Section \ref{grid models}). We therefore explore the effect of different values of the internal temperature, T$_\mathrm{int}$, on the atmosphere and inferred mass fraction of H/He in this planet.

\begin{figure}
    \centering
    \includegraphics[width=1\columnwidth]{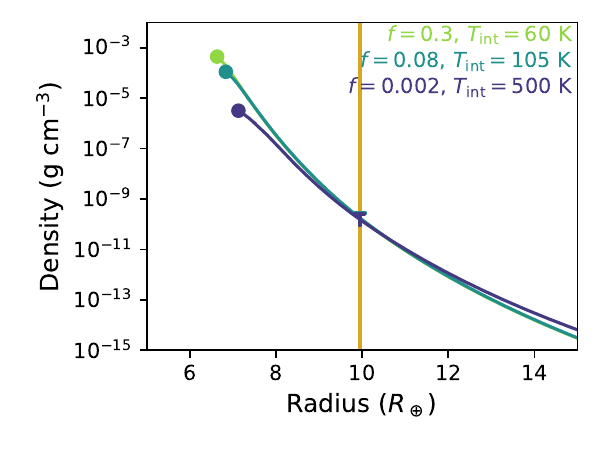}
    \caption{Density-radius profiles of three atmospheric structure models of V1298 Tau b, calculated using the \textsc{aiolos} code \citep{SchulikBooth2023}. See Section \ref{pure H atmo}) for further description of the model. The outer radiative region is shown, with the radiative-convective boundary depicted by dots. Each color line represents a different corresponding atmospheric mass, given as a fraction of the total planet mass $f$: green is $f=0.3$, teal is $f=0.08$, and purple is $f=0.002$. We assume a planet mass of 12~M$_{\oplus}$ from the free retrieval. All three profiles have been simulated assuming the retrieved atmospheric opacity from the observed transmission spectrum (Section \ref{spectrum}). The transit radii (marked with a T and yellow vertical line) is normalized at the TESS optical radius \citep{Feinstein2022} for all three models. However, the three profiles correspond to different internal temperatures of 65, 105, and 500~K and core masses 8~M$_{\oplus}$, 11~M$_{\oplus}$ and 12~M$_{\oplus}$, respectively.}
    \label{fig:profile_comparison}
\end{figure}

In Figure \ref{fig:profile_comparison} we show density profiles in the radiative regions of three atmospheric models, calculated using \textsc{aiolos} \citep{SchulikBooth2023}. We use a mass of 12~M$_{\oplus}$ from the free retrieval and calculate the atmospheric opacity from the best fit PICASO grid model. Each model corresponds to a different total atmospheric mass, expressed as a fraction, $f$, of the planet's total mass, and a different interior temperature. All three have transit radii corresponding to that observed for V1298 Tau b with TESS \citep{Feinstein2022} and match the T-P profile in the radiative region from the best-fit grid model. In the light green is a high-mass atmosphere, $f \sim 0.3$, corresponding to the mass inferred from assuming a fully-convective (i.e. adiabatic) core-envelope structure, such as a \citet{Lopez_2014} model. This profile has the deepest radiative-convective boundary, shown by the dot, and therefore cools slowly: its internal temperature is 65~K at 20~Myr, which is much less than the value inferred from our observations. If we assume the planet has self-consistently evolved for 20~Myr, we find that the planet size can be explained with a lower atmospheric mass fraction ($f=0.08$) and an internal temperature of 105~K yielding the density profile shown in teal. Finally, in dark purple we show the profile corresponding to the inferred $T_\mathrm{int} \sim 500$~K. To be consistent with the radius and inferred high internal temperature from the planet, we find that the planet must have an extended H/He envelope with only a small fraction of the planet's total mass ($\sim$0.2\%). The higher the internal temperature we assume, the more inflated the atmosphere becomes, therefore requiring lower H/He mass fractions to explain the planet size.

\begin{figure*}
    \centering
    \includegraphics[width=1\textwidth]{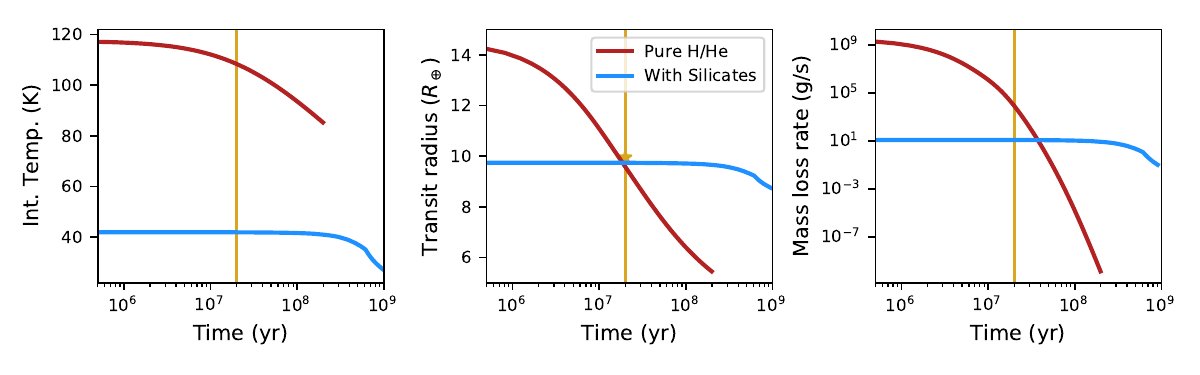}
    \caption{Simulated thermal evolution models for V1298 Tau b for two interior structure scenarios: pure H/He atmosphere (Section \ref{pure H atmo} and an envelop with silicate vapor in equilibrium with a rocky core (Section \ref{silicate atmo}). These models have been calculated using the \texttt{AIOLOS} code \citep{SchulikBooth2023}. The evolution of three key parameters presented: the internal temperature, the transit radii, and the core-powered mass loss rate are presented in the left, middle and right panels respectively. In each panel, the gold line represents the current age of the planet (20~Myr), while the star in the middle panel depicts its observed TESS radius \citep{Feinstein2022}. The red curve is an evolution of a standard, pure H/He atmosphere, using realistic temperature profiles from \textsc{aiolos}. The atmospheric mass fraction is $f=0.08$, with an internal temperature of 105~K. The blue curve represents an evolution with atmospheric structure calculated allowing for chemical equilibrium of silicates and the resulting super-adiabatic temperature profiles. In the pure H/He atmosphere scenario our models predict rapid contraction with V1298 Tau b contracting by $\sim$50\% compared to its current size within the next 100~Myr. In the silicate vapour case the planet does not undergo significant evolution and can maintain its size over Gyr timescale.  }
    \label{fig:evolution_comparison}
\end{figure*}

Each possible atmospheric mass shown in Fig.~\ref{fig:profile_comparison} implies a different evolutionary history, because the cooling timescale depends inversely on the atmospheric mass. For example, if the internal temperature is 500~K, the cooling timescales are extremely short ($<$1~Myr, see Figure \ref{fig:interior_structure_model}), leading to rapid atmospheric contraction. Such rapid contraction in the high internal temperature case makes the planet too small to match the observed radius at the nominal age of the system. On the other hand, self-consistent thermal evolution models using atmospheric profiles produced by the \textsc{aiolos} code predict the internal temperature to be $\sim$105~K. In this case we find that an atmospheric mass fraction of 0.08 matches the observed radius of V1298 Tau b. The evolution in time of this model is shown by the red curve in Fig.~\ref{fig:evolution_comparison}.

In each panel of Fig.~\ref{fig:evolution_comparison} from left to right, we show the evolution in internal temperature, transit radius, and core-powered mass loss rate respectively. In this evolution model, we find the current cooling timescales for V1298 Tau b to be $\sim$10~Myr, comparable to the age of the system. As shown in Fig.~\ref{fig:evolution_comparison}, the model predicts that the planet is currently rapidly contracting due to its short cooling timescale and will decrease in size by nearly 50 percent over the next 100~Myr.

\subsubsection{Scenario 2: silicate vapor envelope in equilibrium with a rocky core } \label{silicate atmo}

The inference of rapid contraction from our evolutionary models can change if we account for recent developments in atmospheric structure theory of sub-Neptunes. Specifically, considering silicate vapor in equilibrium with a rocky core leads to structural changes that slow contraction \citep{misener2022, Misener2023}. To demonstrate this effect, we construct a toy model that includes chemical equilibrium of silicates as described in \citet{Misener2023}. The result of our silicate equilibrium chemistry evolution model with an atmospheric mass fraction  $f=0.08$ is shown in blue in Fig.~\ref{fig:evolution_comparison}. We find that, consistent with previous work, including a non-convective region due to silicate condensation inhibits the contraction of the planet. In this case, the planet can stay at roughly its current radius for hundreds of millions of years. The reason for such slow contraction is two-fold: first, as described in \citet{misener2022}, slight relaxation the super-adiabatic region accommodates much of the interior cooling without significantly changing the atmosphere's extent. Additionally, in order to match the observed radius of V1298 Tau b, in our `With Silicates' case we had to greatly increase the radius of the `atmosphere-interior boundary', which in this toy model is the radius at which $T=4000$~K initially, to 2.45 times the nominal radius of a rocky core used in \citet{Misener2023}. Therefore, the atmosphere is significantly smaller in radial extent compared to the pure H/He case. This compact atmosphere makes the resulting outer atmosphere higher in density, hindering its ability to cool and resulting in much lower internal temperatures, as seen in the left-hand panel. As shown in the right-hand panel of Fig \ref{fig:evolution_comparison}, core-powered mass loss rates remain too low to strip any significant primordial atmosphere on gigayear timescales.

It is worth highlighting that we made significant approximations for the purposes of illustrating the potential effects of silicate condensation at depth on V1298 Tau b. Most importantly, due to the super-adiabatic temperature profiles produced, atmospheric models containing silicate vapor tend to produce planets of small transit radii \citep{misener2022}, leading to our empirical alteration to the `atmosphere-interior boundary' to match the observed transit radius. There are physical reasons to expect the boundary radius to be larger than that of a theoretical pure silicate core: the structural transition between atmosphere and interior in the sub-Neptunes is probably dictated by the miscibility of rock and hydrogen, lying in the solvus between the two materials \citep{young2024}. The evolution of the solvus in sub-Neptunes with time is still uncertain and is the subject of current research in the community. Therefore, although likely qualitatively robust, the quantitative details of the evolution results depicted by the blue curves in Figure~\ref{fig:evolution_comparison} may change as a fully self-consistent treatment of the hydrogen-magma ocean solvus is developed in the future.

\subsubsection{Gas-to-core mass ratio of V1298 Tau b} \label{gcr}

Assuming the mass-radius relationship presented in \citet{Lopez_2014}, we require 30-40\% by the mass gas envelope to explain the measured mass and radius of V1298 Tau b. In this case the expected core mass would be $\sim$ 7~M$_{\oplus}$ and the planet would have to accrete a large amount of gas from the disk, but not enter a run away gas accretion scenario in the core-accretion model \citep{pollack1996}. However, these mass-radius models do not account for the high internal temperature of V1298 Tau b. In Section \ref{pure H atmo} we discuss self-consistent evolutionary models which account for inflation due to the internal temperature. Evolutionary models which predict a 100~K internal temperature for the mass and age of V1298 Tau b require 8\% by mass H/He envelope, implying a core mass of $\sim$11~M$_{\oplus}$. If we consider 500~K internal temperature for V1298 Tau b, the envelope mass fraction is very small (0.2\%) and the core effectively carries the entire mass of the planet (12~M$_{\oplus}$)

Typically, in the core-accretion picture, first a core is formed through pebble/planetesimal accretion. After reaching a mass of few earth masses, it can start accreting H/He from the disk \citep{pollack1996}. As the GCR becomes 1:1 the planet enters a runaway gas accretion phase, resulting in gas giants.  If inflation due to internal heating is considered the inferred core mass is comparable to10~M$_{\oplus}$, a mass limit often assumed for runaway gas accretion. In this case, the inferred GCR is $\sim$10\% and 0.2\% for T$_{int}$=100 and 500~K respectively, implying that the planet did not reach runaway gas accretion. 

The inferred core mass for V1298 Tau b is consistent with core mass estimated for WASP-107b (11$\pm$3~M$_{\oplus}$, \citet{sing2024}). However, the inferred GCR for V1298 Tau b is lower than WASP-107b. Low GCR imply that the accreting core was born in-situ but at a late stage of the disk's life, and thus ran out of its gas reservoir before entering runaway gas accretion \citep{lee_chiang2016}. This is consistent with the relatively low C/O ratio inferred for this planet (Section \ref{low_co_insitu}).  There are other proposed mechanisms, such as disk-gas recirculation \citep{moldenhauer2021} or accretion in a high opacity (dusty) environment \citep{lee2014} which could slow down the gas accretion rate. All of these mechanisms lead to high entropy or `hot start' initial conditions, but given the rapid cooling timescale ($\sim$1000 yr) for such a high internal temperature these mechanisms are unlikely to account for the current internal temperature of this planet. Cold start scenario is also possible for sub-Neptune mass planets: in the `boil-off' scenario as the disk is dissipated the envelope expands rapidly leading to enhanced EUV driven mass loss reducing the H/He mass fraction drastically \citep{owen2016,owen_2020}. However, in this case the internal entropy is significantly lowered right after formation and is inconsistent with the high internal temperature we infer.

\subsubsection{Atmospheric mass loss due to core-powered evaporation} \label{corepowerevaporation}

Since atmospheric mass-loss rates for hydrodynamic escape models depend sensitively on the temperature structure of the upper atmosphere \citep[e.g.][]{MisenerSchulik2024}, we can use the relatively low temperature in the outer atmosphere to constrain the possible future fates of the planet. We use \textsc{aiolos} simulations \citep{SchulikBooth2023} to constrain the mass loss rates due to core-powered mass loss \citep{ginzburg2018,gupta2019}. For the pure H/He envelope scenario (atmospheric mass fraction of 0.08) and assuming a current age of $\sim$20~Myr, corresponding to the teal curve in Fig.~\ref{fig:profile_comparison}, we find from our planetary evolution models a mass-loss rate of $\lesssim$10$^5$~g/s today (see the right panel in Fig.~\ref{fig:evolution_comparison}), although the initial mass-loss rate on 1~Myr timescales may have been as high as $\lesssim$10$^9$~g/s early in the planet's evolution.  If we consider silicates, the calculated mass loss rates at the current age is 10~g/s due to core powered mass loss. Thus, the inferred mass loss from the core-powered mass loss is too low to significantly alter the atmospheric mass over the lifetime of this planet. V1298 Tau b will therefore remain a Sub-Neptune and not evolve into a super-Earth even over Gyr timescales from core-powered mass loss.

\subsubsection{Atmospheric mass loss due to photoevaporation} \label{photoevaporation}

 We calculate the atmospheric mass loss rate due to photoevaporation for the three cases presented in Figure \ref{fig:profile_comparison}. We use the energy-limited approximation \citep[e.g.][]{owen2019}, accounting for the difference between the UV and optical absorption radii, which is important for inflated planets such as V1298 Tau b. The mass loss rate is given by:

 \begin{equation} \label{eq:mdot}
     \dot{M}=\frac{{\eta}R_{\mathrm{UV}}^{2}R_{\mathrm{p}}L_{\mathrm{UV}}}{4a^{2}GM_{\mathrm{p}}}
 \end{equation}

 where $\eta$, the mass loss efficiency, is assumed to be 0.1, the UV luminosity $L_{\mathrm{UV}}$ is assumed to be 10$^{30}$erg s$^{-1}$ \citep{poppenhaeger_21}, and the optical radius $R_\mathrm{p}$ is assumed to be $9.85 R_{\oplus}$ \citep{Feinstein2022}. To calculate the UV radius, we determine where the optical depth to incident UV photons is unity, assuming a UV opacity of 3.6$\times$10$^6$~cm$^2$g$^{-1}$. We find $R_{\mathrm{UV}} \sim 16 R_{\oplus}$ from the density profiles shown in Figure \ref{fig:profile_comparison}. 

 Using this formalism we calculate an instantaneous mass loss rate of 5$\times$10$^{11}$~g\ s$^{-1}$, which implies current mass loss timescales of 7.5~Myr, 320~Myr and 1.2~Gyr for f$_{\mathrm{H/He}}$=0.002, 0.08 and 0.3 respectively. Assuming this constant mass loss rate and the disk dispersal time to be $\sim$10~Myr we calculate that V1298 Tau b has already lost 0.002~M$_{\oplus}$ of its envelope. These mass loss timescales are estimated assuming the current mass loss rate, which does not account for the expected evolution of the stellar luminosity and planet radius with time. 

 In the first case the mass loss timescale is shorter than the estimated age of the system. This would imply that V1298 Tau b is currently in a transient phase and would soon lose its entire primordial atmosphere, ending up as a bare rocky core. However, in this case, the mass of V1298 Tau b does not change appreciably and it ends up as $\sim 12 M_{\oplus}$ rocky super-Earth at the end of the early evolutionary phase. Assuming M-R relations for super-Earths \citep{otegi2020}, a 12~M$_{\oplus}$ rocky planet would have a radius of 1.86~R$_{\oplus}$, resulting in V1298 Tau b ending its evolution in the middle of the sparsely populated radius valley. Therefore, such a low H/He mass fraction at the current age seems unlikely. For the other two cases, the planet could still retain a part of its primordial envelope as the mass loss timescales are significantly longer than 100~Myr, the typical timescale for evolution of the stellar XUV flux.

\subsection{Why is the deep atmosphere of V1298 Tau b so hot?}

The low abundance of methane requires high temperatures in the deep layers of the atmosphere, such that the T-P profile crosses the CH$_4$/CO equality line \citep{fortney_2020}. Subsequent vertical mixing then dredges up the methane poor gas from the interior to the upper observable atmosphere. In this Section we discuss some potential causes for the inferred high temperature in the inner atmosphere.

\subsubsection{Internal energy of the planet}

Assuming a fully convective interior, we can constrain the internal temperature (heat flux from the planet) of V1298 Tau b ($\sim$500~K), which can push the T-P profile to high temperatures in the deep atmospheric layers (Section \ref{grid models}). In Figure \ref{fig:interior_structure_model} (left panel), we show thermal evolution models for a sub-Neptune mass planet. At the current age of this planet (20~Myr), the expected interior temperature is estimated to be $\sim$100K from formation and evolution models. However, we find much higher interior temperature, which would be expected at a much younger age, right after the emergence of the planet from the disk. But the cooling timescale at such high internal temperature would be much shorter than the inside-out dispersal timescale of the disk making this an improbable scenario.

On the other hand, the thermal evolution of the planet could also be slowed if the core has a longer cooling timescale compared to the envelope \citep{vazan2018}, and core cooling timescales longer than the age of the system could sustain the high internal temperature. Another potential mechanism to increase the temperature in the inner atmosphere would be to have silicate vapour in the deep atmosphere in chemical equilibrium with the rocky core \citep{misener2022}, or silicate rainout which may occur during the first $\sim$100~Myr \citep{vazan2024}.

\subsubsection{Energy from dynamical interactions} \label{heating mechnisms}

The internal temperature of V1298 Tau b could be explained by external heating mechanisms, which would replenish the energy lost to thermal cooling. A high internal temperature, similar to V1298 Tau b has also been reported for WASP-107b \citep{welbanks2024, sing2024}. Tidal heating has been proposed as a potential solution to explain the large temperature. The eccentricity of V1298 Tau b is not well constrained, however ongoing TTV studies hint toward relatively low eccentricity ($\sim$0.01, Livingston et al. in prep.). However, given the youth of this system, tidal effects cannot be ruled out. It could have tidal synchronization timescales $\sim$100~Myr \citep{Seager2002},  longer than its age . It is also possible that the planetary spin is misaligned with its orbital angular momentum. \citet{millholand_2020} showed that spin axis obliquity could increase tidal heating by a large factor. Applying the formalism described in \citet{millholand_2020}, we estimate the interior temperature of V1298 Tau b as a function of the tidal quality factor (Q) and the obliquity of the spin axis. We find that we require high spin obliquity ($>$80$^{\circ}$) and low tidal quality factor ($\sim$100) to obtain internal temperatures higher than 400~K. This tidal quality factor is comparable to Neptune (Q=170, \citet{trafton1974}). The estimation of internal temperature incorporating tidal heating is shown in Appendix \ref{appendix:tidal luminodity calc}. 

If the internal temperature of V1298 Tau b is 500~K it would result in a luminosity of 1.7$\times$10$^{27}$~ergs/s. The gravitational binding energy of the planet is 3$\times$10$^{42}$~ergs. It implies that if the luminosity is due to dynamical interactions, it could lead to a significant orbital evolution of this system with a timescale of $\sim$50~Myr. This would imply that the V1298 Tau system would be dynamically unstable

 Population synthesis studies \citep{mordasini2018,lambrechts2019} show that sub-Neptunes and super-Earths are formed in multi-planet systems in mean-motion resonance, similar to V1298 Tau system. However, after disk dispersal dynamical interactions between planets significantly perturb their orbits and erase the initial orbital architecture over 10-100Myr after disk dispersal \citep{izidoro2021,marimbu2024}. These perturbations lead to breaking of the resonant chains. Such dynamical interactions could potentially convert some of the orbital energy of these planets to internal energy.

Given the youth of V1298 Tau system and its aligned geometry \citep{gaidos2022,marshall2022}, it is possible that we are witnessing ongoing dynamical interactions between the planets in this system. Recent studies have shown that a large fraction of the young planets discovered are in multi-transiting systems. Most of these systems are found near MMR, unlike mature sub-Neptune/super-Earth systems \citep{dai2024}. This hints towards the fact that young transiting planet systems ($<$100Myr) are dynamically cold and are likely to undergo/currently undergoing a phase of dynamical interactions with each other. Collisions with remnant debris from planet formation could also supply energy to the young V1298 Tau planets. Such remnants (exocomets) have been found around another young system, $\beta$ Pictoris \citep{des_etangs2022}.

\subsubsection{Inefficient heat transport due to compositional gradients} \label{double diffusion convection}

In the last couple of sub-sections we discuss the possible mechanisms to produce a large amount of energy in the planet's interior, such that it can push the T-P profile to very high values. However, it is possible to achieve high temperatures in the deep layers without high heat flux from the planet's interior if the heat transport is inefficient. 

Typically, a uniform composition is assumed in gas envelopes, which leads to overturning convection in the interior, resulting in efficient heat transport, cooling, and an adiabatic interior. In this case high interior temperature corresponds to higher temperature in the deep atmosphere. However, metallicity gradients can hamper large scale convection in the interior, leading to non-adiabatic interior. Under certain conditions, heat can be trapped in the deep atmosphere, leading to high atmospheric temperatures at these layers (non-adiabatic), even for low internal temperatures. This effect has been discussed in the context of gas giants \citep{chabrier2007,leconte2012}. Condensation-inhibited convection, higher up in a saturated atmosphere, may be another mechanism to slow the planet's cooling and keep the deep atmosphere hot \citep{guillot1995,leconte2017}.

Atmospheric metallicity gradients are a natural result of the formation of planets by pebble accretion \citep{ormel2021,bloot2023}. A schematic diagram is shown in Figure \ref{fig:interior_structure_model} (right panel), which explains the gradually mixed interior structure. The low luminosity of Uranus \citep{nettleman2013} can be explained using the metallicity gradient structure \citep{vazan2020} which leads to high temperature in the deep atmosphere due to heat trapping. V1298 Tau b could be similar to Uranus in this context; it could be born with a metallicity gradient, which it still retains at the current age. This would lead to high temperature in the deep atmosphere without the need for high internal luminosity, thereby reconciling V1298 Tau b with early evolution models (Figure \ref{fig:interior_structure_model}). The presence of silicate vapor in the deep atmosphere could also lead to non-adiabatic interior thermal structure as discussed in Section \ref{silicate atmo}.

\subsection{V1298 Tau b in comparison with mature sub-Neptunes/super-Earths} \label{sub-N population}

 We find $\sim10\times$ solar atmospheric metallicity and sub-solar C/O ratio. V1298 Tau b is a sub-Neptune or possibly even a super-Earth progenitor, but with an atmospheric metallicity comparable to gas giants \citep[e.g,][]{rustomkulov2023,xue2024,gagnebin2024,August2023}. 

\begin{figure}
    \centering
    \includegraphics[width=\linewidth]{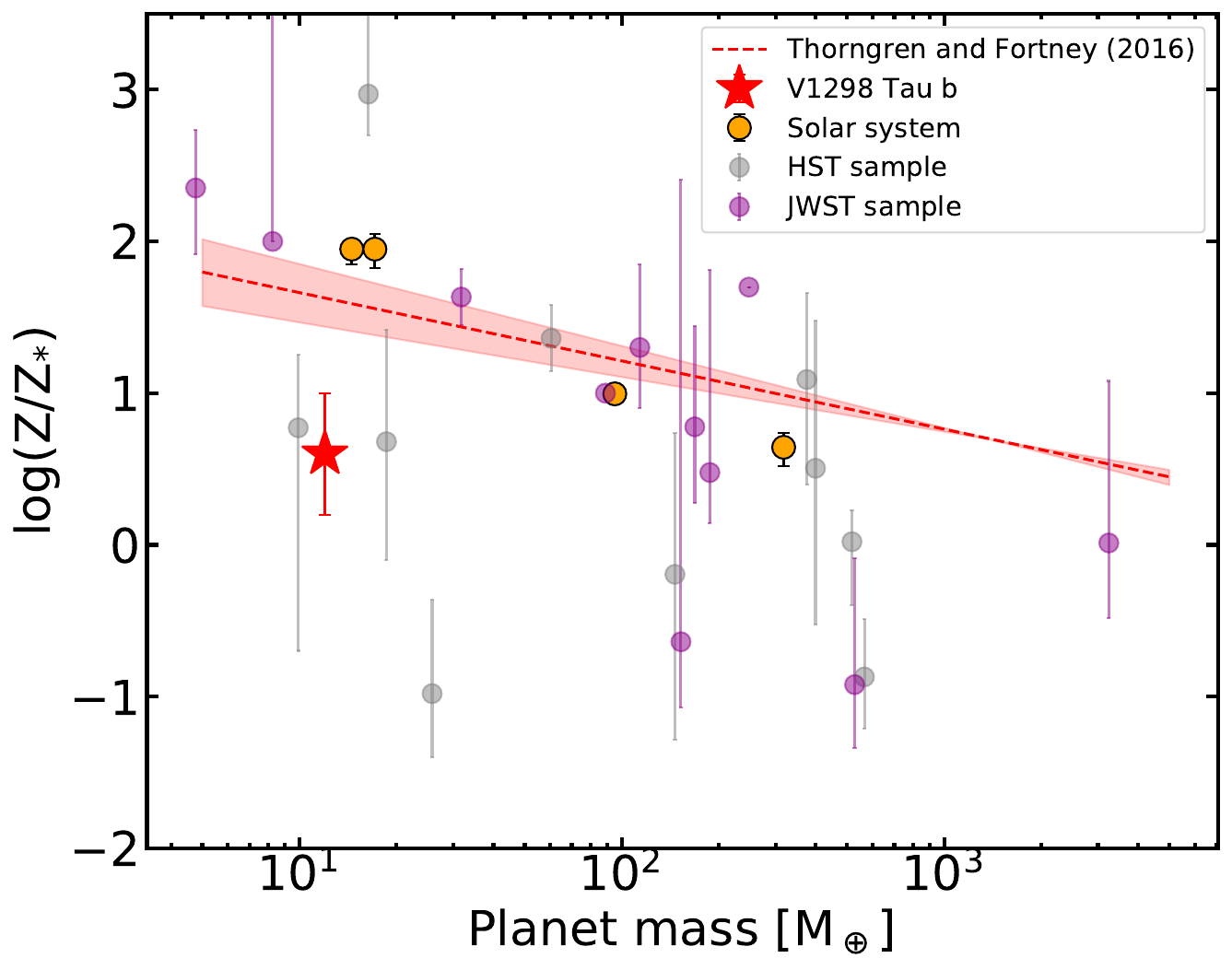}
    \caption{Mass-atmospheric metallicity diagram showing V1298 Tau b (red star with 1$\sigma$ uncertainty) in the context of the solar system planets (orange points), and the exoplanet population with published atmospheric metallicity. We assume the mass and atmospheric metallicity of V1298 Tau b from the free chemistry atmospheric retrieval (Appendix \ref{appendix: free retrieval}, Table \ref{tab:retrieval table}). The retrieved atmospheric metallicity is converted into stellar units assuming a stellar metallicity of 0.10$\pm$0.15 dex \citep{mascareno_2021} and the uncertainties were propagated. We show exoplanets with metallicity measurements from HST in grey, taken from \citet{wakeford_dalba} and with JWST (purple points). The JWST sample has been taken from Table I1. of \citet{barat2024b}. The red dashed line and the shaded region show the mass-metallicity relation derived in \citet{thorngren2016}. The atmospheric metallicity of V1298 Tau b is lower than what is expected from core-accretion formation models \citep{pollack1996,thorngren2016} and what is typically seen for known sub-Neptune/super-Earth planets by more than one order of magnitude, indicating ongoing evolutionary mechanisms which are expected to enhance the atmospheric metallicity of V1298 tau b as it matures. }
    \label{fig:mass-metallicity}
\end{figure}

A mass-metallicity trend (decreasing metallicity for increasing mass) has been predicted for the core-accretion formation scenario \citep{pollack1996,boddenheimer1986} and has been found in the solar system, as well as in exoplanets \citep{welbanks2019,thorngren2016}. In Figure \ref{fig:mass-metallicity} we show the mass-atmospheric metallicity diagram, with V1298 Tau b using the mass and metallicity constraints we derive, the solar system planets and exoplanets with atmospheric metallicity measured using HST and/or JWST. The atmospheric metallicity constraint of V1298 Tau b puts it $\sim$5$\sigma$ away from the exoplanet mass-metallicity trend \citep{thorngren2019}, and an order of magnitude lower compared to the solar system ice giants as well as mature sub-Neptunes.

V1298 Tau b has a clear and relatively metal-poor atmosphere, distinct from its mature sub-Neptune/super-Earth counterparts. Sub-Neptunes/super-Earth with temperatures similar to V1298 Tau b have been found to host haze-dominated/metal-rich atmospheres, such as GJ1214b \citep{desert2011a,kreidberg14,kempton2023,gao2023}, TOI-836c \citep{wallack2024} and Kepler-51 planets \citep{libby_robert2020}. GJ3470b, has a mass and equilibrium temperature (11~M$_\oplus$, 600K) similar to V1298 Tau b, and shows similar molecular features of H$_2$O, CH$_4$, SO$_2$ and CO$_2$ \citep{beaty2024, benneke_19}, but with higher atmospheric metallicity ($\sim$100$\times$solar) and the presence of hazes. Therefore, the absence of haze and the relatively low metallicity atmosphere of V1298 Tau b make it distinct compared to its mature counterparts. 

HIP-67522b \citep{rizutto2020} is another young transiting planet (17~Myr old), which was observed by JWST in cycle 1. It is similar in size compared to V1298 Tau b. Interestingly, its transmission spectrum has also revealed a relatively clear atmosphere with a large scale height \citep{thao2024}, similar to V1298 Tau b. Its transmission spectrum also constrains its mass in the sub-Neptune regime ($\sim$14~M$_{\oplus}$), comparable to V1298 Tau b. Interestingly, its atmospheric metallicity is also relatively low ($\sim$30$\times$solar) compared to mature sub-Neptunes. Thus, the similarity between the atmospheres of these two young transiting planets indicates that young sub-Neptunes (atleast some of them) are born with low metallicity envelopes and undergo early evolutionary processes with increasing their atmospheric metallicity with age.

 In pebble accretion theory, a metallicity gradient is formed naturally in the envelope \citep{ormel2021,bloot2023}. In Figure \ref{fig:interior_structure_model} (right panel) we show a possible gradually mixed structure for V1298 Tau b and a schematic diagram to explain this and compare with the standard core-envelope picture. Moreover, interiors containing compositional gradients well explain the measurements of giant planets in the solar system \citep{wahl2017,2mankovich2021,vazan2020}. At the current age, both the gradually mixed envelope and core-envelope structures produce similar metallicity for the upper atmosphere for this planet. However, as illustrated in the schematic diagram in Figure \ref{fig:interior_structure_model}, in the gradually mixed envelope structure, atmospheric mass loss during early evolution could remove the metal-poor upper layers of the atmosphere, revealing the more metal-rich inner layers of this planet. It is possible that given its young age, we are witnessing the primordial envelope of a sub-Neptune/super-Earth progenitor, which was born with a composition gradient, and has not had the time for overturning convection to homogenize the atmospheric composition.

Early evolutionary processes such as mass loss can lead to enhanced atmospheric metallicity due to preferential loss of lighter elements. This `metal-drag' effect is specifically important for low mass sub-Neptunes with low envelope mass fraction, where metallicity could be enhanced by one order of magnitude due to mass loss \citep{Malsky2020}. This process does not significantly affect the atmospheric metallicity of the gas giants \citep{louca2023}. But, for sub-Neptunes, atmospheric mass loss could lead to significant preferential loss of lighter elements (H/He) and lead to metallicity enhancement by an order of magnitude \citep{louca2025}. This model is consistent with the current atmospheric metallicity of V1298 Tau b which through preferential loss of lighter elements could gain an order of magnitude more metals in its atmosphere as it matures. 

Another mechanism for inducing preferntial mass loss for sub-Neptunes would be to have metallicity gradients in its envelope. Metallicity gradients in the atmosphere could lead to preferential loss of metal-poor gas from the upper atmosphere, thereby revealing the metal-rich inner layers. This scenario can explain the apparent low atmospheric metallicity of V1298 Tau b (and HIP 67522b), and coupled with early evolutionary mechanisms can reconcile its metal-poor atmosphere with mature sub-Neptunes/super-Earths. As the atmospheric metallicity increases, hazes may also start emerging in the atmosphere of V1298 Tau b. Interestingly, atmospheric metallicity gradients can also lead to heat trapping and explain the high temperatures needed in the deep atmosphere to explain the low methane abundance, without the need for extremely high internal temperatures (Section \ref{double diffusion convection}).

\section{Summary}

We present the JWST/NIRSpec G395H transmission spectrum of V1298 Tau b (20-30~Myr old). We combine the JWST spectrum with previously published HST/WFC3 spectrum of the same planet. We find an offset of $\sim$400~ppm between the observed transit depths at the different epochs, which is comparable to the predicted offset from the difference in optical brightness of the star at the two epochs.

From the combined JWST and HST spectra (Figure \ref{fig:gas contribution}) we find a mostly clear and extended atmosphere with a large scale height ($\sim$1500~km). Efficient haze formation can be ruled out (Figure \ref{fig:spectra_full}). We detect CO$_2$ (35$\sigma$), H$_2$O (30$\sigma$), CO (10$\sigma$) CH$_4$ (6$\sigma$), SO$_2$ (4$\sigma$) and OCS (3.5$\sigma$) from free atmospheric chemistry retrieval using PICASO. We also fit self-consistent forward models grids including disequilibrium chemistry (vertical mixing and photochemistry) generated using PICASO and ATMO to the observed spectrum. The constraints on the atmospheric parameters from the PICASO and ATMO grids are consistent with each other (Table \ref{tab:retrieval table}). The retrieved molecular abundances from the PICASO free chemistry retrieval are consistent with the abundances from the best fit PICASO grid model for all molecular species except SO$_2$, which is underpredicted by the grid (Figure \ref{fig:comparison_free_grid}).

We measure the mass of V1298 Tau b directly from the transmission spectrum. The PICASO free retrieval reports a mass of 12$\pm$1~M$_{\oplus}$ and the PICASO and ATMO grids report 15$\pm$1.7 and 15~M$_{\oplus}$ respectively. It confirms V1298 Tau b as a sub-Neptune and rules out one previous RV mass measurement reporting a Jovian mass \citep{mascareno_2021} at 40$\sigma$ confidence level, but is consistent with RV mass upper limits from \citet{sikora2023,finoceti2023}. The mass derived from the transmission spectrum is consistent within 1$\sigma$ to the TTV mass measurement for this planet (Livingston et al. in prep). The mass measurement along with the large scale for V1298 Tau b is consistent with the gas dwarf scenario for sub-Neptunes.

We derive an atmospheric metallicity and C/O ratio from the PICASO free chemistry retrieval of log(Z/Z$_{\odot}$)=0.6$^{+0.4}_{-0.6}$ and 0.22$\pm$0.06 respectively. These constraints are consistent with constraints we derive from the PICASO grids (logZ/$Z_{\odot}$=1.05$\pm$0.2, C/O=0.23$\pm$0.08) within  1$\sigma$ (Figure \ref{fig:metal_co_grid_free_comparison}). The best-fit metallicity from the ATMO grids is consistent within 1$\sigma$ to the PICASO grid and free retrieval. But the C/O ratio from ATMO is higher than the PICASO results which is likely to be due to differences between the two grids with regards to how C/O ratio is changed.  The retrieved atmospheric metallicity of V1298 Tau b is low compared to mature sub-Neptunes/super-Earths (by atleast an order of magnitude) which are known to have metal-rich ($>$100$\times$solar) haze dominated atmospheres \citep{desert2011a,morley2013,bean2021,kempton2023,madhusudhan2023,benneke2024,beaty2024}, thereby challenging the mass-metallicity prediction from core-accretion formation model \citep{pollack1996,thorngren2019}.

 We find a methane VMR of log[CH$_4$]=-6.2$^{+0.3}_{-0.5}$ from the PICASO free retrieval, which is $\sim$ 2 orders of magnitude (7$\sigma$) lower than the equilibrium chemistry prediction. The best fit PICASO and ATMO grids require extremely high internal temperature (500 and 600~K respectively) along with vertical mixing (K$_{zz}$=10$^7$ and 10$^8$ respectively) to explain the low methane abundance. 

 We calculate internal structure models including the effect of the internal temperature. We consider two models: 100~K internal temperature (self-consistent early evolution) and 500~K (retrieved from spectrum). Our models find that we require a low gas-to-core ratio or GCR (10 and 0.1\% respectively) with massive cores (11 and 12~M$_{\oplus}$ respectively) to explain the mass and radius of V1298 Tau b. Although our models find that core-powered mass loss is low for this planet, photoevaporation could remove a large portion of its primordial envelope. In the first case, the mass loss timescale is $\sim$300~Myr and if the stellar activity subsides by then, V1298 Tau b ends up as a sub-Neptune with a thin envelope ($<$1\%). But in the 500~K case, it loses all of its primordial envelope and ends up as a rocky core within the next 7.5~Myr. The low GCR inferred for V1298 Tau b is consistent with formation within the ice line, either in a dissipating disk \citep{lee_chiang2016}, dusty environment \citep{lee2014} or with disk-gas recirculation \citep{moldenhauer2021}, which would prevent the massive core from reaching runaway gas accretion. Furthermore, the sub-solar C/O ratio is also consistent with in situ formation within the water ice line.

Such a high internal temperature is inconsistent with conventional early evolutionary models, which predict $\sim$100-150~K for a planet mass between 10-15M$_{\oplus}$ at its current age (Figure \ref{fig:interior_structure_model}). The cooling timescale for such a high internal temperature is short ($<$1~Myr), and therefore to sustain such a high internal temperature we require other heating mechanisms, such as tidal heating or dynamical interactions between multiple planets in the V1298 Tau system. To explain the internal temperature through tidal heating, we require a low tidal quality factor ($\sim$ 100) and a high obliquity between the spin and orbital angular momentum of the planet ($>80^{\circ}$).

Alternatively, V1298 Tau b could have a slower cooling than predicted. The presence of silicate vapor \citep{misener2022} or a primordial metallicity gradient \citep{ormel2021,bloot2023} in the deep atmosphere could make the thermal structure non-adiabatic and increase the cooling timescale of the planet. Composition gradients hamper large scale convection and lead to non-adiabatic interior \citep{chabrier2007,leconte2012,leconte2017}. If the interior is non-convective, the temperature of the deep atmosphere could be significantly hotter compared to the fully convective case without the necessity of extremely high internal heat flux from the planet. 

In the metallicity gradient scenario, atmospheric mass loss processes could remove the upper layers of the atmosphere that are metal-poor and increase the metallicity of the observable atmosphere with age \citep{fortney2013,vazan2022}. This evolutionary process of metallicity enrichment could reconcile the metal-poor atmosphere of V1298 Tau b with mature sub-Neptunes. Thus, atmospheric metallicity gradients could explain both the high internal temperature and the low atmospheric metallicity of V1298 Tau b. The similarly aged HIP-67522b (17~Myr) also appears to have a relatively low metallicity compared to its mass \citep{thao2024} and is consistent with the atmospheric composition evolution scenario. Further observations of young transiting planets between 10-100~Myr are needed to confirm if a metallicity-age trend exists, and what is the evolution timescale.

J.M.D acknowledges support from the Amsterdam Academic Alliance (AAA)
Program, and the European Research Council (ERC) European Union’s Horizon 2020 research and innovation program (grant agreement no. 679633;
Exo-Atmos). This work is part of the research program VIDI New Frontiers in
Exoplanetary Climatology with project number 614.001.601, which is (partly)
financed by the Dutch Research Council (NWO). Y.K. acknowledges support from JSPS KAKENHI Grant Numbers 21K13984, 22H05150, and 23H01224, and the Special Postdoctoral Researcher Program at RIKEN.
G.W.H. acknowledges long-term support for our automatic telescopes from NASA, NSF, and Tennessee State University.
A.V. acknowledges support by ISF grants 770/21 and 773/21. A.D.F. is supported by NASA through the NASA Hubble Fellowship grant HST-HF2-51530.001-A, awarded by the Space Telescope Science Institute, which is operated by the Association of Universities for Research in Astronomy, Inc., for NASA, under contract NAS 5-26555. J.H.L. acknowledges support from JSPS KAKENHI Grant Numbers 24H00017 and 25K17450.

\newpage
\bibliography{references.bib}{}
\bibliographystyle{aasjournal}
\appendix
\restartappendixnumbering

\section{Eureka! data reduction} \label{eureka}

In Stage 1 of the data reduction, we converted the up-the-ramp counts to a slope. We applied a group-level background correction by subtracting column-wise median counts for each group of an integration. This procedure has been noted to correct for the 1/f noise seen in the JWST detectors \citep{Schlawin2020,alderson2023}. We find that including this step at the Stage 1 of the data reduction improves the SNR of the final light curves by a factor of $\sim$2. We mask the trace while estimating the background in this step. A spatial cosmic ray filter is applied at this stage with a rejection threshold of 5$\sigma$. We tested the impact of the threshold by performing data reduction with 4-7$\sigma$ threshold, however, this exercise did not significantly impact the SNR in the final light curves.

In Stage 3 of the data reduction we perform an optimal extraction \citep{horne1986} from the reduced images to obtain 1D stellar spectra. We correct for the curvature of the trace in both NRS1 and NRS2. We perform another column-by-column median background subtraction at this stage. We used different aperture widths (5-10 pixels) to assess its impact. We find that 8-pixel aperture results in best SNR for the final light curves, therefore we use 8 pixels for the final optimal extraction window. The rest of the pixels in the column are used to estimate the background. The optimal extraction is performed using the `meddata' option in \texttt{Eureka!}. A median of all frames is constructed, which is smoothed in the spectral direction using a median filter. A 5$\sigma$ rejection threshold is used while constructing the median profile. The smoothed median profile is used to derive weights for optimal extraction. A similar approach has been used in other data sets analyzed with \texttt{Eureka!} (e.g, see \citet{May2023,benneke2024}).

By comparing the y-position of the trace for each integration we conclude that the telescope pointing was stable within 0.01 pixels throughout the visit. We determined the x-position drift by cross-correlating the 1D extracted spectra of each integration with the first integration. We find that the RMS x-drift is 0.03 pixels. 

\section{SPARTA data reduction} \label{sparta}

Starting from \texttt{uncal.fit}, \texttt{SPARTA} provides completely independent data reduction without using routines from any other existing pipelines. Here we provide a summary of the \texttt{SPARTA} pipeline, focusing on where it differs from \citet{kempton2023} and \citet{Zhang2024}:

\begin{enumerate}
    \item \texttt{calibrate.py}: the first step is reference pixel correction. We use the top and bottom 4 pixels as the reference pixels and subtract the median value of all odd/even columns of these pixels within a given group, then proceed to linearity correction. Following that, we introduce a group-level column-by-column and row-by-row background subtraction, as follows:

    First, we choose the brightest pixel in each column as the trace's center. To remove outliers, we smooth it using \texttt{scipy.ndimage.uniformfilter} at size = 100. Then, to get more precise trace center positions, we fit a cubic polynomial to it as a function of the column number. Pixels that are $>$ 8 pixels vertically away from the center are masked. For each column, we subtract the median value of the unmasked part as the background. Similarly, we perform row-by-row background subtraction by subtracting the median of the part that is horizontally $>$ 750 pixels away from the center. 

    Next, we apply up-the-ramp fitting, gain correction, and dark correction in the same way as \citet{Zhang2024}.

    \item \texttt{remove\_bkd.py} repeats the column-by-column and row-by-row background subtraction on an integration level using the same approach described above. 

    \item \texttt{get\_positions\_and\_median\_image.py} calculates the position shifts of the trace along both spectral and dispersion directions and then obtains the median image across all integrations.

    \item \texttt{optimal\_extraction.py} uses the median image and the positions to calculate the optimal profile. Then we extracted the spectra with a half-window size of 13 pixels.

    \item \texttt{gather\_and\_filter.py} first rejects 3$\sigma$ outliers of the smoothed binned light curve obtained with \texttt{scipy.ndimage.median\_filter} with a filter size of 50. Then we manually removed columns [409:420], [785:803], [228:230] for NRS1, and [320:348], [1200:1258], [691:704] for NRS2. Finally, we repair the light curve by rejecting 3$\sigma$ outliers of the smoothed unbinned spectroscopic light curves. We then extract and generate the light curves in the desired wavelength ranges.

\end{enumerate}
\begin{figure}
    \centering
    \includegraphics[width=0.8\linewidth]{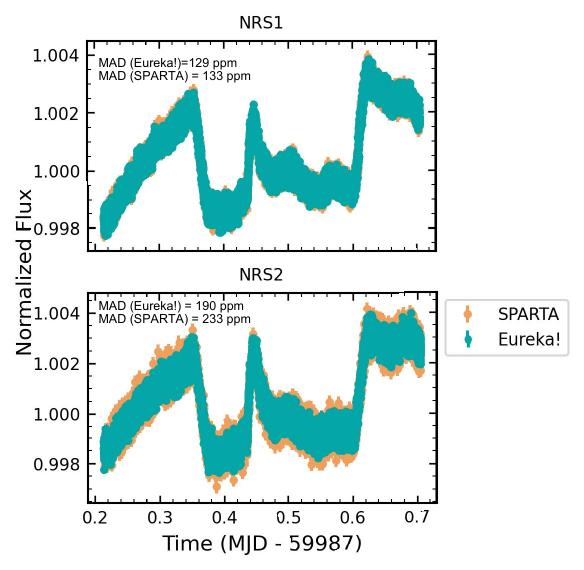}    \caption{Comparison between white light curves in NRS1 (upper panel) and NRS2 (lower panel) obtain from \texttt{Eureka!} data reduction (Appendix \ref{eureka}) and SPARTA data reduction (Appendix \ref{sparta}) The light curves and scatter are comparable in NRS1 (129 and 133 ppm for Eureka and SPARTA respectively). However, in NRS2 SPARTA shows a comparatively higher scatter (190 and 233 ppm for Eureka and SPARTA respectively). Both reductions show qualitatively similar light curve features: a low frequency temporal variability signal due to the rotational modulation of spots on the photosphere of the rapidly rotating (2.8 days) young T-Tauri star, and high frequency signal in the middle of the transit which is due to a flare and prominent post flare oscillations which continue even after egress, confirming that this periodic variability signal is due to the star and not spot crossings by the planet.}
    \label{fig:eureka_sparta_comparison}
\end{figure}

\section{Fitting JWST light curves} \label{subsec:lightcurve_analysis}

\subsection{Empirical light curve model fit} \label{empirical model fit}

Using the Stage 3 output from \texttt{Eureka!} (Appendix \ref{eureka}) we produced the JWST white transit light curve of V1298 Tau b. We use custom tools to extract the light curves. For the NRS1 and NRS2 white-light curves we consider wavelength ranges of 2.78-3.72~$\mu$m and 3.8-5.12~$\mu$m respectively. We apply a sliding filter outlier rejection algorithm to the light curves with a threshold of 10$\sigma$.

The transit light curve shows significant deviations from what is expected for quiet stars observed with this instrument and at this fluence \citep{alderson2023}. Two main time dependent deviations can be observed at different frequencies which we interpret as due to stellar variability. The low frequency variations are caused by the rotation of the heterogeneous photospheric surface as expected for this T-Tauri star \citep{david2019b,Feinstein2022}. At higher frequency, an abrupt change in flux emerges during the transit that is identifiable as the photometric signature of a flare.

Assuming the stellar photosphere and the spots are blackbodies, the ratio of their Planck functions can be used to convert the optical variability to any other wavelength \citep{czesla2009,desert2011b}.
We converted the photometric R-band variability observed from the ground (Section \ref{ground based data description}) to HST and JWST bandpasses. We calculate a conversion factor following the formalism outlined in \citet{desert2011b} to HST and JWST wavelengths, assuming a 5000~K photosphere and a 4000~K spot temperature \citep{biagini2024}.  The scaled variability in the HST and JWST bandpasses are shown (Figure \ref{fig:v1298_photometry}), where we also show the median subtracted variability observed during the HST and JWST visit. The long-term variability seen over the course of the entire observation is consistent with rotational variability of the T-Tauri star, which has a period of 2.8 days \citep{david2019b} . This level of variability is expected from previous observations from K2 \citep{david2019b}, TESS \citep{Feinstein2022} as well.

The flare is seen during the transit and subsequent post-flare oscillations are also apparent (See Figure \ref{fig:white_lc}). We can see these oscillations even after egress,  further confirming that these variations are indeed due to a flaring event and not due to spot crossings. To continue with the light curve analysis, we first need to remove the flare from the transit light curve, and we chose to employ and test both empirical and data-driven methods.

Spectroscopic light curves are derived using 10 pixel bins for both NRS1 and NRS2 resulting in 150 and 200 channels respectively. The spectroscopic light curves are derived using the same procedure as the white-light curves from \texttt{Eureka!} 1D stellar spectra.

We first model both the white and the spectroscopic light curves using an empirical mathematical function given in Equation \ref{eq:eq1} to simultaneously fit for the long-term variability, flare, post-flare oscillations and the transit, where t is the time from the start of the visit, P(t) is the transit model, simulated using \texttt{batman}. We fix all orbital parameters, except transit depth, semi-major axis, mid-transit time, and linear limb darkening coefficient for the white-light curves. We fit a linear limb darkening coefficient. F(t) is the stellar flare and oscillation model. A$_{0}$ is the flare amplitude, t$_{0}$ is the mid-flare time, ${\tau}_{rise}$ and ${\tau}_{fall}$ are the rise and fall timescales of the flare. A$_{1}$ is the amplitude of the sinusoidal perturbation.  We fit the white-light curve using an MCMC \citep{emcee} with 50 walkers and 50000 steps. The first 5000 steps are considered as burn-in phase and discarded. 

\begin{equation} 
    M(t)=P(t) \times C_{1}\times(1+C_{2}t+C_{3}t^{2}+F(t))
\end{equation}

\begin{equation} \label{eq:eq1}
    F(t)=
    \begin{cases}
    A_{0}e^{\frac{t-t_{0}}{{\tau}_{rise}}} & \text{if }   t<t_{0} \\
    A_{0}e^{\frac{t_{0}-t}{{\tau}_{fall}}}+A_{1}\sin(\omega (t-t_{0})+\phi) & \text{if }   t>t_{0} \\
\end{cases}
\end{equation}

We use a quadratic polynomial function for the baseline,  We tested higher order polynomial functions, however we found the second-order polynomial function to provide the most favorable Bayesian Information Criteria (BIC) value. To model the flare, we assumed a double exponential function with different rise and decay timescales and a sinusoid to model the post-flare oscillations. We also test a different flare model: gaussian rise and exponential decay profile \citep{pitkin2014} to estimate the effect of the assumed flare profile. A double exponential model is preferred over a gaussian rise profile by a $\Delta$BIC of $\sim$2000 for the NRS1 white light curve and $\sim$1000 for the NRS2 white light curve.

\begin{figure}
    \centering
    \includegraphics[width=0.5\linewidth]
    {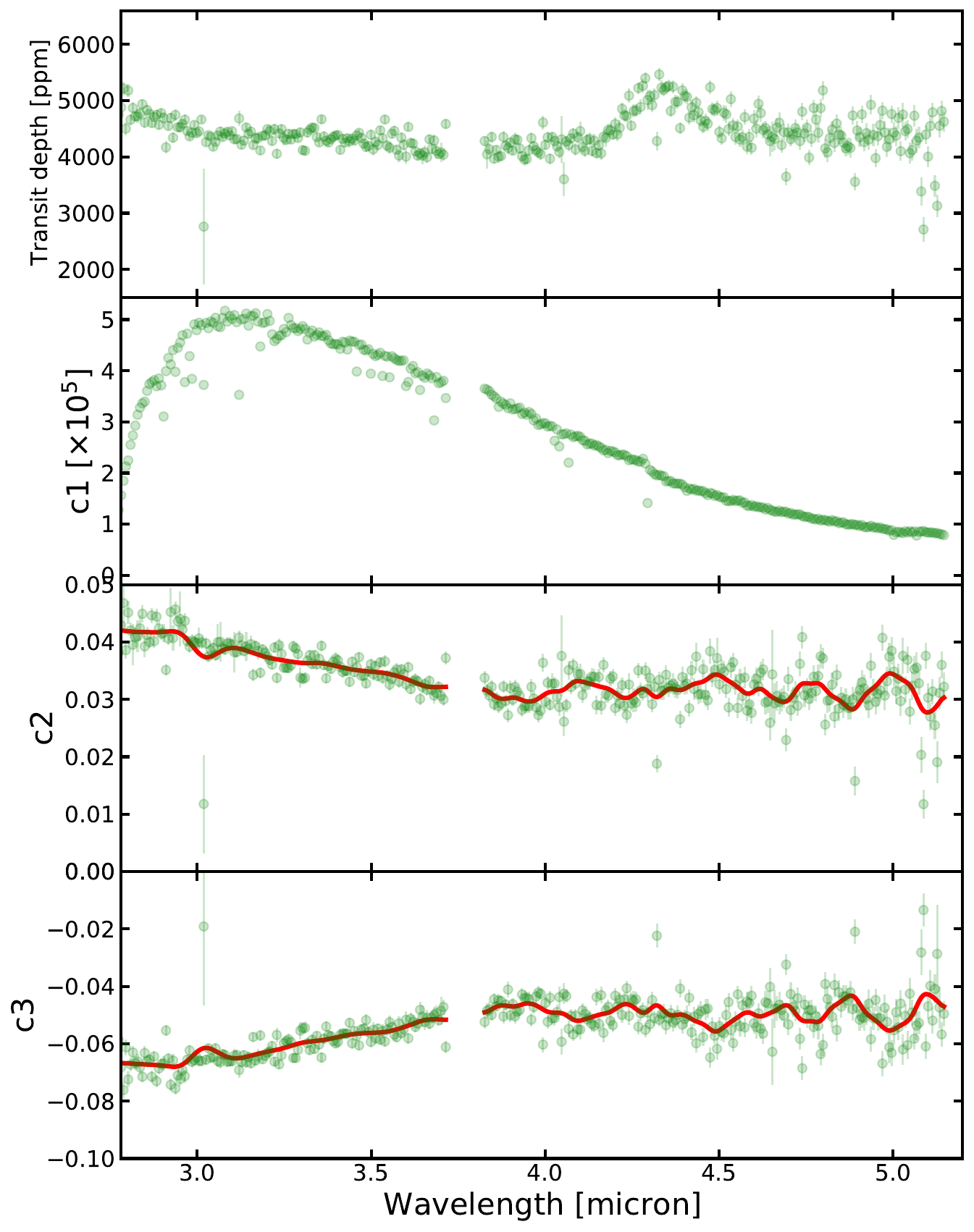}
    \caption{Transmission spectrum of V1298 Tau b (upper panel) and fitted baseline quadratic polynomial coefficients (middle and lower panels), defined in Equation \ref{eq:eq1}. This transmission spectrum has been derived from empirical model fit (Appendix \ref{subsec:lightcurve_analysis}) to spectroscopic light curves derived from 10 pixel bins on the 1D stellar spectra from \texttt{Eureka!} data reduction (Appendix \ref{eureka}). The light curve models are described in Eqn \ref{eq:eq1}. The model includes a quadratic polynomial function to model the low frequency stellar baseline variability, a double exponential function to model the flare, a sinusoidal function to model the post-flare oscillations and a \texttt{batman} model for the planet transit. The solid red lines show a smoothed function, obtained by applying a gaussian filter on the observed c$_2$(linear coefficient) and c$_3$ (quadratic coefficient) values. The transmission spectrum derived using this approach exhibits prominent H$_2$O, CH$_4$ and CO$_2$ absorption features, but shows significant scatter and multiple outliers ($>$5$\sigma$ compared to neighbouring data points).  We find that c$_2$ and  c$_3$ appear correlated with the transit depth: outliers in the transmission spectrum correspond to outliers in the c$_2$ and c$_3$. c$_2$ and c$_3$ show a strong wavelength dependence for NRS1, and are relatively wavelength independent for NRS2. The smoothed c$_2$ and c$_3$ coefficients are used for the data-driven detrending of the spectroscopic light curves in the second iteration of light curve fits described in Appendix \ref{subsec:lightcurve_analysis}.}
    \label{figure:polynomial_spectrum_correlation}
\end{figure}

For the spectroscopic light curves we follow an identical approach, but we fix the mid-transit time and semi-major axis from the respective white light curve fits. We fit for the linear limb darkening coefficient. We fit for the polynomial coefficients, flare amplitude, rise and decay timescale terms. We also fit for the amplitude, frequency, and phase of the sinusoid. The planetary transit is modeled using \texttt{batman} \citep{batman}. 
The transmission spectrum derived from this approach is shown in Figure \ref{figure:polynomial_spectrum_correlation}.

We found that the baseline polynomial coefficients are correlated with the spectroscopic transit depths (Figure \ref{figure:polynomial_spectrum_correlation}), and also significant residuals in the white-light curve (Figure \ref{fig:white_lc}). This indicates that the empirical model is not fully able to model the astrophysical systematics.  Outliers in the transmission spectrum also appear as outliers in the polynomial coefficient spectra. Since the baseline model can be curved, we find that sometimes the baseline can `fill-in' a part of the transit, resulting in shallower measured transit depths. This effect is an artifact of the polynomial model we assume. As c$_2$ and c$_3$ are correlated to the transit depth, fitting all three together introduces noise in the transmission spectrum. The polynomial coefficients are themselves anticorrelated. Therefore, the polynomial is a necessary evil: the young star with rotational variability demands higher order polynomials to model its curved baseline, however, the polynomial function increases the uncertainty on the transit depth and can significantly increase the scatter. 

A strong wavelength dependence for c$_2$ and c$_3$ is seen in Figure \ref{figure:polynomial_spectrum_correlation} for NRS1. The baseline is a combination of stellar variability and instrumental systematics. Visit long slopes have previously been reported from the JWST commissioning data \citep{espinoza2023}, which show significant wavelength dependence for NRS1 and relatively constant values for NRS2. The wavelength dependence seen in the NRS1 polynomial coefficients and the relative wavelength independence in NRS2 could be due to a detector level effect, as argued in \citet{espinoza2023}.

\subsection{Data-driven light curve fitting technique} \label{data-driven detrending}

The`data-driven' approach for detrending the observed spectroscopic light curves bypasses the need to fit for the polynomial baseline and in principle can account for any residual systematics in the light curves which could not be accounted for using the empirical flare/sinusoid models. We follow up our first spectroscopic light curve analysis with a data-driven approach to detrend the spectroscopic light curves. We smooth the polynomial coefficients shown in Figure \ref{figure:polynomial_spectrum_correlation} with a gaussian kernel. Using the smoothed coefficients, we derive wavelength-dependent long-term variability profiles to detrend the low frequency rotational variability in the spectroscopic light curves. We assume that rotational variability will not lead to abrupt changes in polynomial coefficients as a function of wavelength, and outliers seen in Figure \ref{figure:polynomial_spectrum_correlation} are artifacts of the light curve fitting process.

\begin{figure*}
    \includegraphics[width=1\linewidth]{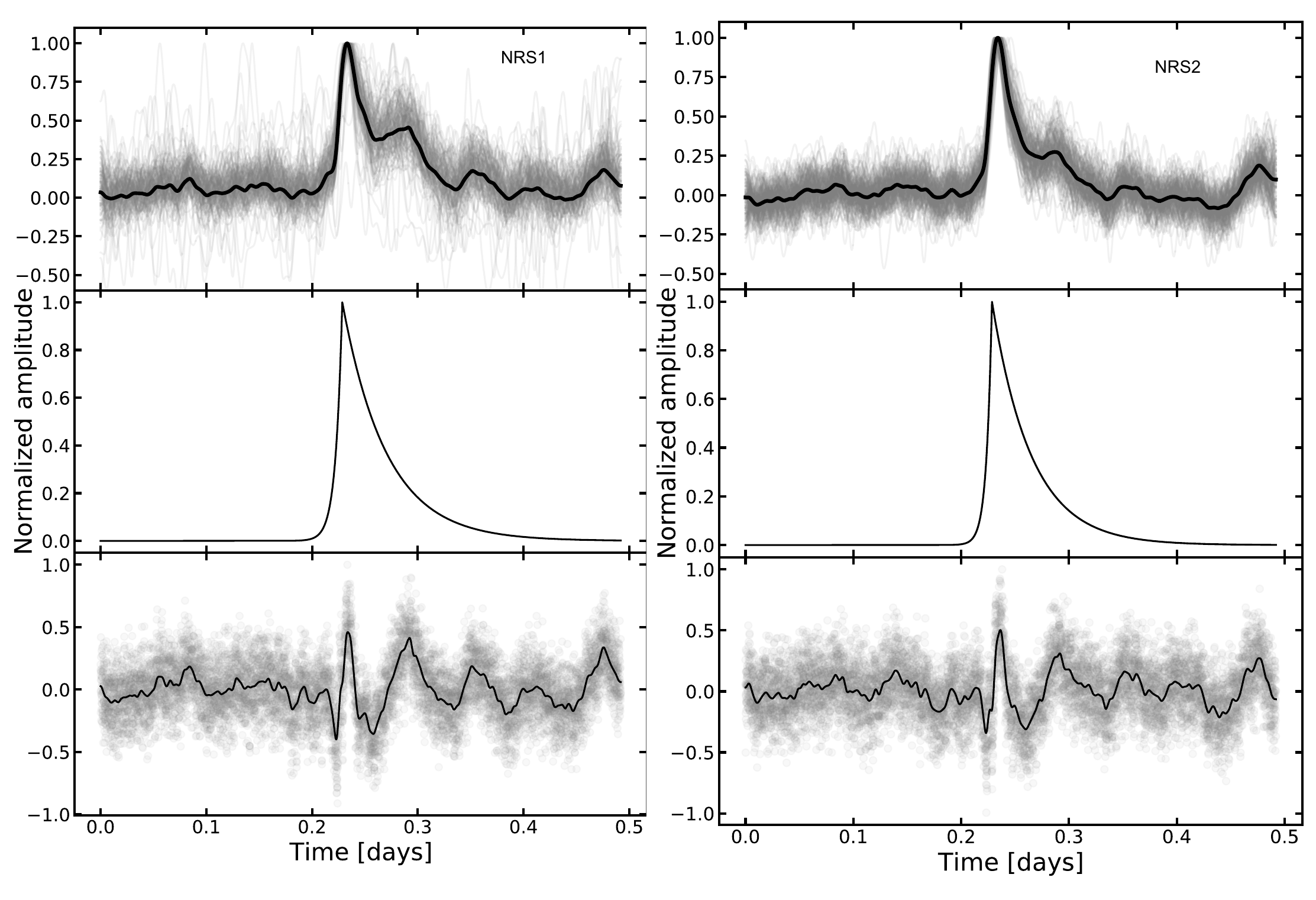}
    \caption{Upper panel: Spectroscopic light curves of V1298 Tau b derived from the \texttt{Eureka!} data reduction (Appendix \ref{eureka}) with the best fit planetary transit and polynomial baseline removed (grey lines). These light curves have been derived by fitting the spectroscopic light curves with an empirical model (Eqn C2) described in Section \ref{subsec:lightcurve_analysis}.  Subsequently the observed spectroscopic light curves were divided by the best fit planetary transit model and the quadratic baseline model, effectively leaving behind the astrophysical systematics (flare and oscillations). The light curves have been normalized at the peak of the flare for visual comparison. The light curves for all spectroscopic channels show similar temporal behavior, i.e we do not see any wavelength dependent temporal stretching. The solid black lines shows the median of the individual flare only light curves. Middle panel: Best fit flare model from white light curve fits. The flare model is derived from Eqn \ref{eq:eq1} using the best fit flare parameters from the white light curve fits. This profile has been used in a second iteration fit to the spectroscopic light curves as described in Appendix \ref{data-driven detrending}) Lower panel: Residual oscillation profile from white light curve fit. The residual oscillations are derived from the white light curves by removing the transit, baseline and the best fit flare model (middle panel). The solid black line shown in the lower panels has been derived by applying a gaussian filter to the observed residual oscillation (grey points). The median residual oscillation profile has been used to detrend the the spectroscopic light curves by multiplying with a scaling factor in a second iteration data-driven fit as discussed in Appendix \ref{data-driven detrending}.}
    \label{fig:flare_profile}
\end{figure*}

In Figure \ref{fig:flare_profile} (upper panel) we show the flare-oscillation profiles derived from the spectroscopic light curves by removing the best fit transit and baseline polynomial from the empirical model fit to the light curves. These residual profiles show similar temporal behavior. This means that the flare and oscillations show similar timescales (rise and decay timescales and oscillation frequency) for all the wavelength channels. Therefore, in the data-driven approach, we model the flare and the post-flare oscillations using the flare profile derived from the white-light curve (Figure \ref{fig:flare_profile}, middle and lower panel). We apply independent scaling factors for the flare and residual oscillation profiles. In this approach, we remain agnostic about the exact mathematical form of the oscillation term. Therefore, in the data-driven approach, we fit for the planet transit a linear baseline, an amplitude for the flare, an amplitude for the the residual oscillation profile, and a linear limb darkening law. The semi-major axis and mid-transit time are fixed from the empirical model fit to the white-light curves.

The transmission spectrum we derive from the data-driven spectroscopic light curve fits is shown in Figure \ref{fig:gas contribution}. The data-driven approach results in transit depth uncertainties within 5\% of the expected photon noise limit for all spectroscopic channels. In Figure \ref{fig:detrended_lc} we show stacked detrended spectroscopic light curves obtained using the data-driven technique. It shows that the low frequency variability signal has been removed from the spectroscopic light curves. However, four horizontal streaks are seen across the stacked detrended light curves. Upon visual inspection of the light curves at these channels we find that the temporal shape of the flare (i.e., the shape of the flare and oscillation light curves) are significantly different compared to the white-light curves in these channels.  As a result our data-driven detrending method, which assumes that the flare-oscillation profiles show similar temporal behavior across wavelength, does not perform well for these channels. The reduced chi-square for all the spectroscopic light curves in the data-driven fitting approach is comparable to 1. However, for the four channels that appear as outliers in Figure \ref{fig:detrended_lc}, the reduced chi square is greater than 10 and thus we eliminate these channels from the rest of our analysis.  We find that the four channels which appear as outliers in the transmission spectrum correspond to the cores of stellar lines: 3.03~$\mu$m (HI Pfund $\epsilon$), 3.29~$\mu$m (HI Pfund $\delta$), 4.04~$\mu$m (HI Brackett $\alpha$) and 4.65~$\mu$m (HI Pfund $\beta$).

\section{Free atmospheric chemistry retrieval} \label{appendix: free retrieval}

We perform an isothermal $T(P)$ profile retrieval with a total of 17 free parameters including temperature $T(P)=T$, gray cloud opacity $\kappa_{\rm cld}$, mass of the planet $M_{\rm p}$, the 1 mbar radius of the planet, and two offset parameters -- one between the observed transmission spectra with HST/WFC3 and NIRSpec/G395H NRS1 detector and another between the observations of NIRSpec/G395H NRS1 and NIRSpec/G395H NRS2. In addition to these, we include the volume mixing ratios of H$_2$O, CO, CO$_2$, CH$_4$, NH$_3$, SO$_2$, H$_2$S, OCS, HCN, and N$_2$ as free parameters as well. These abundances have been assumed to be constant with altitude/pressure in our free chemistry retrieval setup. The 1 mbar radius of the planet has been included as a free parameter assuming $R(\text{1 mbar})$=$R_{\rm p}$(1+$X_R$), where $R_{\rm p}$ is 0.85 $R_{\rm jup}$ and $X_R$ is the free parameter to allow for a slight alteration of the 1 mbar radius. We have assumed uniform priors for all of these 16 free parameters. The mass of the planet has been allowed to vary between 0.0314558 $M_{\rm jup}$ and 0.0943673 $M_{\rm jup}$, which correspond to 10-30 Earth masses. We use the PyMultiNest Nested sampling package to perform our retrievals with 2000 live points. The offset parameter between the HST/WFC3 and NIRSpec/G395H spectra has been allowed to vary between 0 and 600 ppm added to the HST/WFC3 spectrum.

In order to compute the detection significances of CH$_4$, SO$_2$, and CO, we take a conservative approach of performing three additional retrievals on the transmission spectrum by setting the abundance of CH$_4$ once, SO$_2$, CO and OCS once to 0. We compare the Bayesian evidence of these additional retrievals with the Bayesian evidence obtained by the full retrieval to compute the detection significance of these two gases. The detection significance of CH$_4$, SO$_2$, CO and OCS are found to be 6$\sigma$, 4$\sigma$, 10$\sigma$, and 3.5 $\sigma$ respectively.

\begin{figure}
    \centering
    \includegraphics[width=1\linewidth]{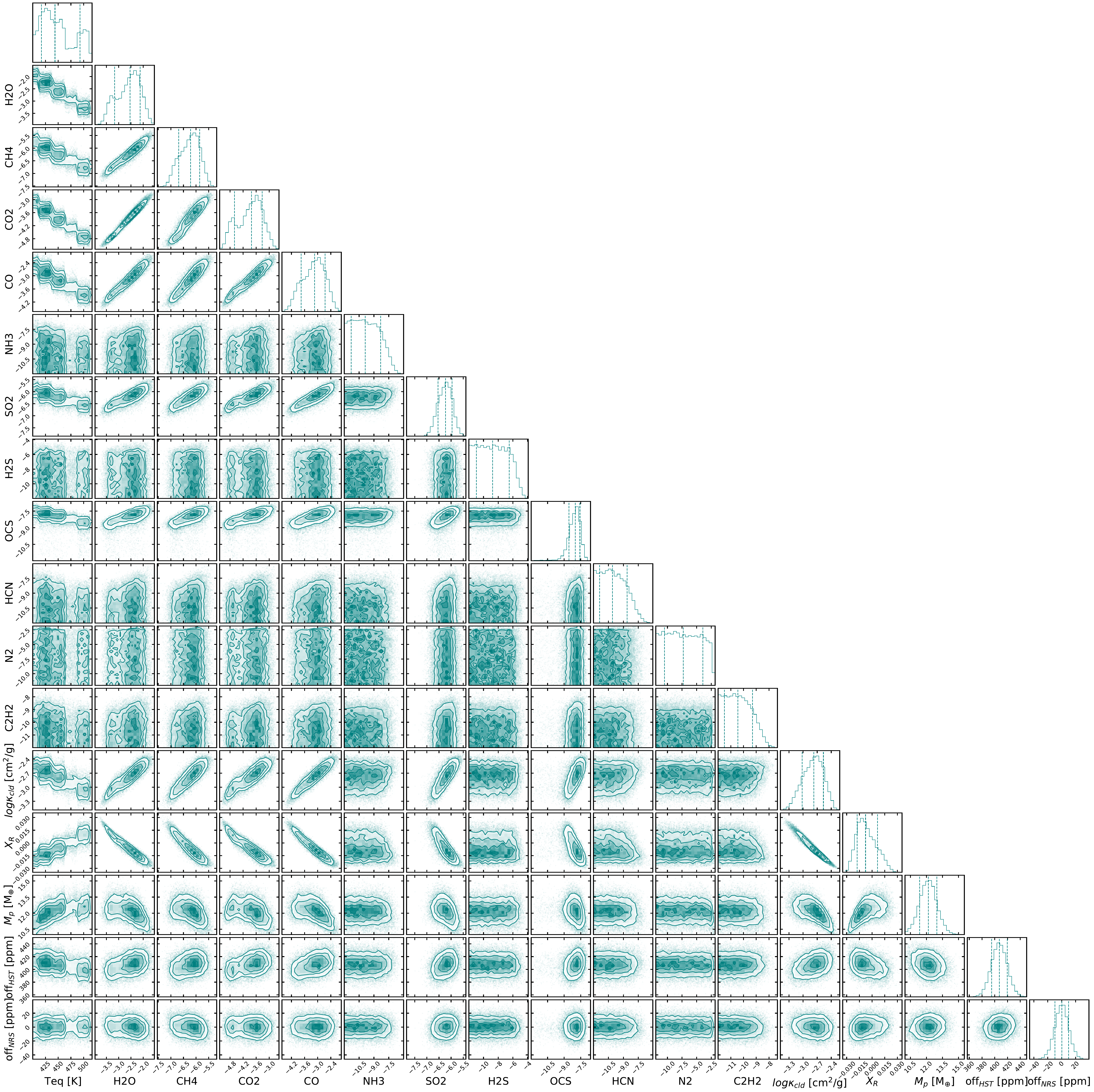}
    \caption{Posterior distribution from free atmospheric chemistry retrieval with PICASO on the combined HST and JWST transmission spectrum of V1298 Tau b. See Section \ref{spectrum} and Appendix \ref{appendix: free retrieval} for further details. We included the mass as a free parameter in these retrievals and find 12$\pm$1M$_{\oplus}$. The free retrieval finds H$_2$O, CH$_4$, CO$_2$ and CO. The free retrievals also find OCS and SO$_2$. We retrieve upper limits for other molecules included in our model. We do not find significant offset between NRS1 and NRS2 however we find $\sim$400 ppm offset between HST and JWST.}
    \label{fig:v1298 free chemistry corner}
\end{figure}

\section{PICASO Forward models for V1298 Tau b} \label{appendix: forward models}

{The grid of forward models was computed assuming a smaller heat redistribution parameter than in the case of full redistribution of incident stellar energy between the day and night side of the planet. This was done so that the photosphere of the RCTE models has temperatures closer to the retrieved isothermal temperature in the free retrieval. We compute several $T(P)$ profiles for the planet varying $T_{\rm int}$ between 100 K to 600 K with an interval of 50 K. The C/O of the atmosphere is assumed to be 0.458 and the atmospheric metallicity is varied between [M/H]=+0.0 and [M/H]=+2.0 while computing these $T(P)$ profiles for the planet. The stellar properties used to compute this grid of forward models are $T_{\rm eff}$= =4970 K, $log(g)$ =4.25, [M/H] =-0.1,
$R_{\rm star}$ = 1.31 R$_{\rm \odot}$, a = 0.1688 AU. Even though the mass of the planet is quite uncertain, we assume a planet mass of 0.0754938 $M_{\rm jup}$ and a planet radius of 0.85 $R_{\rm jup}$ for computing these forward models. We incorporate the mass uncertainty of the planet in our analysis with a step described further below.}

{The RCTE $T(P)$ profiles are used as inputs in the \texttt{photochem} 1D chemical kinetics model \citep{wogan23} to model disequilibrium chemistry due to atmospheric vertical mixing and photochemistry in the planet's atmosphere. To model the effect of  photochemistry on the planet's atmospheric chemistry we use the observed host stars high energy SED from \citet{duvvuri23}. Using the pre-calculated $T(P)$ profiles, we generate the model atmospheric chemistry for the planet by varying atmospheric metallicity between +0.3 and +1.95 with an interval of +0.15 dex. For each of these models, the RCTE $T(P)$ profile calculated at the nearest metallicity point is used. We also vary the atmospheric C/O from 0.0229 to 0.687 with an interval of 0.0916. Note that this C/O is reflective of the bulk C/O of the planet as we remove 20\% of the gas-phase O- owing to condensation in the deep atmosphere of the planet. The strength of atmospheric vertical mixing is parametrized with the 1D eddy diffusion parameter $K_{\rm zz}$. The $K_{\rm zz}$ for the atmosphere was assumed to be constant with altitude in each model and was varied between 10$^6$ cm$^2$s$^-1$ and 10$^{10}$ cm$^2$s$^-1$, with an interval of 1 dex in log($K_{\rm zz}$). This variation in $K_{\rm zz}$ reflects the typical uncertainty in our understanding of mixing processes in giant planet and brown dwarf atmospheres \citep{sing2024,welbanks2024,Mukherjee2022a,phillips2020,miles20,mukherjee24}.}

{Variation of all these parameters leads to a large grid of photochemical forward models for V1298 tau b containing 11$\times$12$\times$6$\times$5=3960 unique forward models for the planet's atmosphere ($T_{\rm int}$x[M/H]x(C/O)x$K_{\rm zz}$). We generate the transmission spectrum from these grid of models while varying four additional parameters --  $X_R$, $M_{\rm planet}$, $\kappa_{\rm cld}$, and offset between HST and G395H/NRS1. $X_R$, $M_{\rm planet}$, $\kappa_{\rm cld}$, and the offset is defined exactly the same way as for the free retrieval exercise. The priors for these parameters are also set identically with the free retrieval exercise.} 

{We first use this forward modeling grid to calculate the $\chi^2$ value corresponding to each grid model. This $\chi^2$ value is used to assign the probability of each grid model using $p(x)=exp(-\chi^2/2)$, where $x$ represents the grid point. Using these probabilities as weights, we plot the corner plot that we obtain the grid fit to the data. The grid fit prefers the $T_{\rm int}$ parameter to be between 450-600 K. The preferred metallicity from the grid is between 4-10$\times$solar and the C/O is found to be sub-solar. The grid also prefers a relatively low strength of vertical mixing around $\sim$10$^7$cm$^2$/s. The forward model grids are often inflexible as they are computationally expensive to compute. As a result, the spacing in between grid points are often too large to bring out realistic uncertainties on the grid parameters when fitting observational data. We have chosen to sweep through a larger parameter space to estimate the bulk parameters like {$T_{\rm int}$} instead of computing very finely spaces grid of models in this work.}

\begin{figure}
    \centering
    \includegraphics[width=1\linewidth]{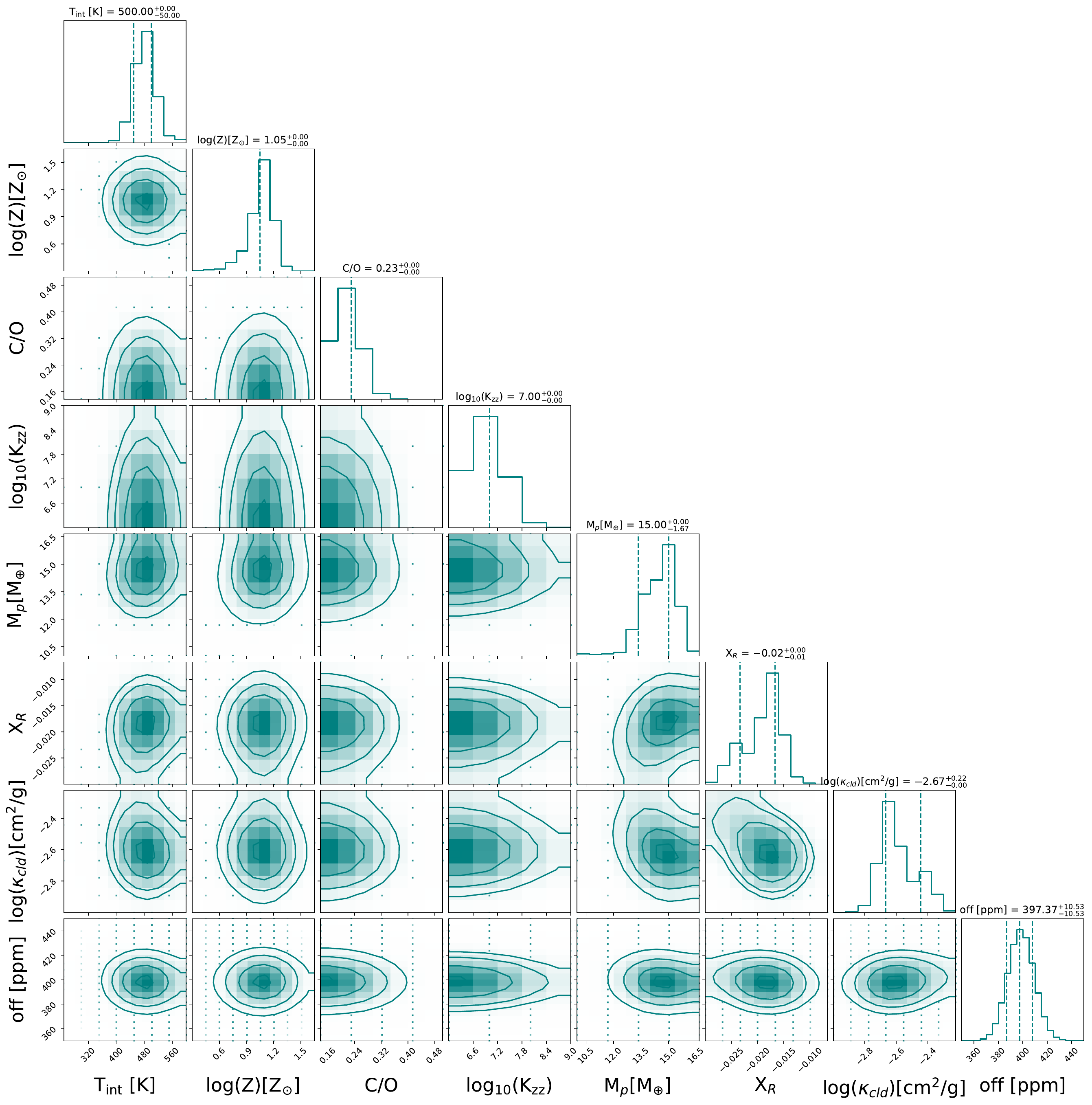}
    \caption{ Posterior distribution from grid fitting based on self-consistent PICASO forward models (Appendix \ref{appendix: forward models}). Each grid model has been assigned a probability $p(x)=exp(-\chi^2/2)$ to compute the posterior distribution. We include mass of the planet as a free parameter too. The best-fit mass (14.67 M$_\oplus$) from the grid is consistent with what we found in the free retrieval.  The offsets retrieved between HST and JWST is also consistent with the free retrieval. The best fit parameters from this grid are tabulated in Table \ref{tab:retrieval table}. }
    \label{fig:grid_corner}
\end{figure}

\begin{figure}
    \centering
    \includegraphics[width=0.75\linewidth]{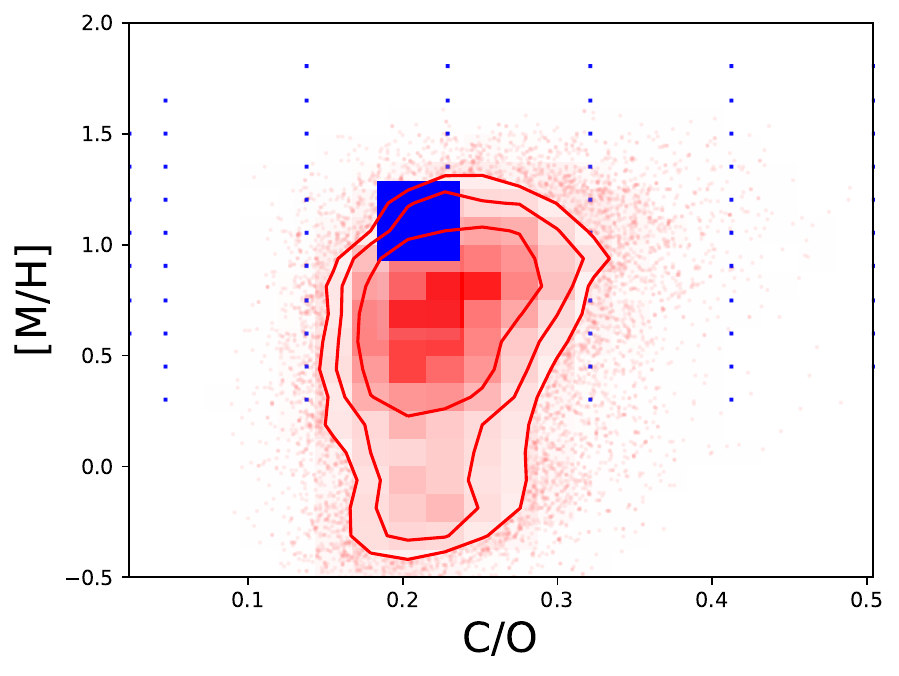}
    \caption{Comparison between metallicity (solar units) and C/O ratio derived from the PICASO free atmospheric chemistry retrieval performed on the combined HST and JWST transmission spectrum of V1298 Tau b (Appendix \ref{appendix: free retrieval}) and the best fit PICASO grid model (Appendix \ref{appendix: forward models}). The red contours are from the free retrieval and the blue square represents the estimate from the grid fit. The contours represent 1$\sigma$ level confidence intervals. The retrieved atmospheric parameters from both the free retrieval and grid fit are given in Table \ref{tab:retrieval table}. The blue dots show the spacing of the grid used (See Appendix \ref{appendix: forward models}). We find that both the metallicities and the C/O ratio are consistent within 1$\sigma$ level of confidence between the free retrieval and the grid. }
    \label{fig:metal_co_grid_free_comparison}
\end{figure}

\section{ATMO self-consistent forward models} \label{appendix:atmo_model_descriptions}

We generate RCTE $P$-$T$ profiles using ATMO, for which we assume the same stellar and orbital properties as used for the PICASO grid. However, in our grid we include day-night recirculation factor as a grid variable, unlike the PICASO grid. Therefore, in the final fit to the observed spectrum we do not include any temperature offset term. The re-circualtion factor used in the ATMO grid is detailed in \citet{arora2024}.

The generated $P$-$T$ profiles are used as inputs for VULCAN \citep{Tsai_2021} to calculate chemical profiles. VULCAN includes disequilibrium chemistry effects such as vertical mixing and photochemistry. In VULCAN, we run all the models with the S-N-C-H-O reaction network for the dis-equilibrium chemistry. The UV stellar flux from \citet{Duvvuri2023} utilized for driving the photo-chemistry in the model ranges from 100 nm to 899 nm. To initialize the mixing ratios of gases for dis-equilibrium chemistry runs, we use volume mixing ratios obtained from the equilibrium chemistry calculations using \textsc{FastChem} \citep{stock2018fastchem}. For the \textsc{FastChem} calculations, the solar elemental abundances of elements present in the reaction network, i.e., S-N-C-H-O along with He, are extracted from ATMO for consistency. For each RCTE $P$-$T$ profile, we scale the elemental abundances with the required metallicity and C/O accordingly. Note that [C/H] is fixed, and [O/H] is varied to account for [C/O]. H$_2$ is considered the dominant gas in the atmosphere. K$_{zz}$ is assumed to be even across the atmosphere layers in the VULCAN simulations. We also note that among all the model simulations in the grid some of the VULCAN models do not converge. This could be attributed to the numerical instabilities in the model. Therefore, we are only able to find the best-fit forward model with these converged models from the ATMO plus VULCAN grid.

\section{Ground based photometric monitoring of V1298 Tau} \label{v1298_photometric_monitoring}

\begin{figure}
    \centering
    \includegraphics[width=1\linewidth]{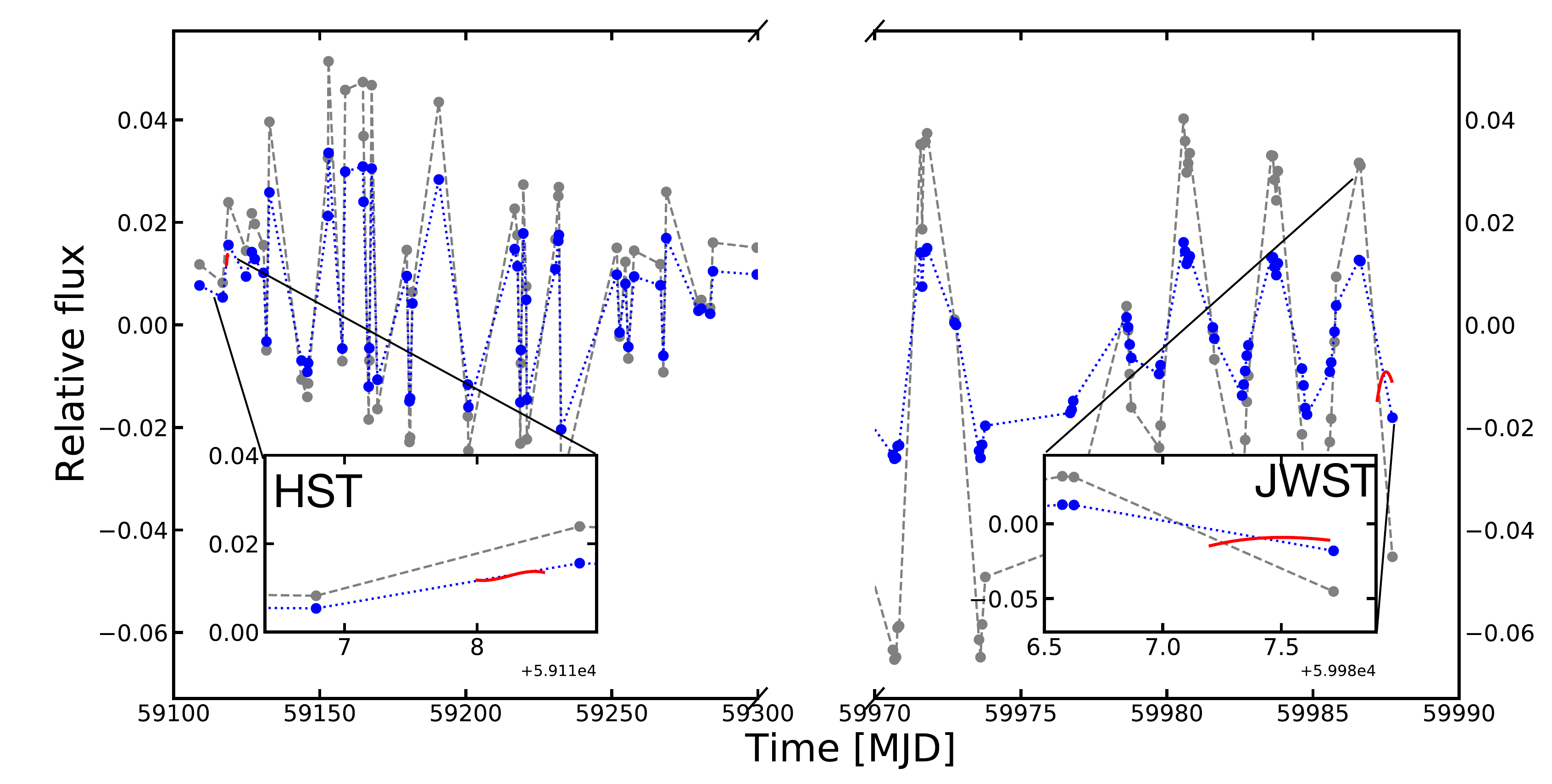}
    \caption{R-band photometric variability of V1298 Tau observed using the ground based Tennessee State University C14 0.36m Automated Imaging Telescope (AIT) at Fairborn Observatory (gray points). See Section \ref{ground based data description} and Appendix \ref{v1298_photometric_monitoring}) for further details of the ground based monitoring observations. We show observations from Sector 2 and 5 from this monitoring campaign which coincided  with the HST (left) and JWST (right) observations. We have assumed the median of the two epochs as the quiescent stellar flux and subtracted it.  We have scaled the R-band variability to the median HST (1.35$\mu$m) and JWST/NIRSpec NRS2 (4.5$\mu$m) wavelengths respectively using formalism outlined in \citet{desert2011b}. We assume a photosphere temperature of 5000~K \citep{david19} and a spot temperature of 4000~K \citep{biagini2024}. The scaled variability in the two bands is shown with a blue dashed line. The solid red lines show the stellar baseline (derived from the respective best fit white light curves) during the HST/JWST observations. These baselines have been median subtracted and vertically offset for visual comparison with the scaled relative variability model.}
    \label{fig:v1298_photometry}
\end{figure}

We acquired 509 observations of V1298 Tau in the Cousins $R$ band with the
TSU Celestron 14-inch (C14) Automated Imaging Telescope (AIT) from late in
the 2019-20 observing season through the 2023-24 season.  The AIT uses an 
SBIG STL-1001E CCD camera with a Kodak KAF-1001E detector that gives a 
21 x 21 arcmin field of view.  Each nightly observation consisted of 3-5 
consecutive exposures centered on V1298 Tau.  The individual nightly frames 
were co-added and reduced to differential magnitudes in the sense V1298 Tau 
minus the mean brightness of four constant comparison stars in the same 
field of view.  Each nightly observation has been corrected for bias, 
ﬂat-ﬁelding, pier-side offset, and for differential atmospheric extinction. The uncertainty in the nightly observations is typically around to 0.0015 to  0.0030 mag, as determined from observations of the constant comparison stars. Further details on the operation of the C14 and the analysis of the data can be found in \citet{sing2015}.

 In Figure \ref{fig:v1298_photometry} we show the photometric observations of V1298 Tau around the HST (18th October 2020) and JWST (12th February 2023) epochs. We convert the R-band variability to HST/WFC3 and JWST/NIRSpec bands using the formalism outlined in \citet{desert2011b}. We assume a spot temperature of 4000~K and a photospheric temperature of 5000~K for V1298 Tau. Assuming that the stellar variability scales as the ratio of the two blackbody functions \citep{Knutson2009}, we can convert the R-band variability to any different wavelength. In Figure \ref{fig:v1298_photometry} we show the scaled variability of the HST (left) and the JWST NRS2 (right) bandpasses. Although the photometry cadence is low, the variability we see for both the HST and JWST observations qualitatively agree with the predicted variability from the optical photometry assuming spots to be the major source of this variability.

We have found a large offset between the HST and JWST white light curve transit depths ($\sim$400~ppm). We find that V1298 Tau was brighter by 0.08 mag in the R-band on the night of the HST observations compared to the JWST visit. Difference in baseline flux level is known to affect the measured transit depth \citep{czesla2009}. Using the formalism described above, we convert the difference in R-magnitude to the HST bandpass. If we account for stellar flux variability between the different observation epochs (HST and JWST) using the formalism outlined in \citet{desert2011b}, we find that the white light curve transit depth in the HST bandpass would increase by 380~ppm compared to what has been measured \citep{barat2023}. Thus, the observed offset between the JWST and HST observations is comparable to the expected offset from the difference in optical brightness of the star between the two epochs.

\section{Details of microphysical haze model} \label{appendix: haze details}

 We first performed photochemical simulations to derive the distribution of gaseous species. Then, since the production rate of haze monomers is uncertain for exoplanets, we assumed a certain fraction, $f_\mathrm{haze}$, of the sum of the photodissociation rates of the precursor molecules would result in haze monomer production. Here, we define the precursor molecules as $\mathrm{CH_4}$, $\mathrm{HCN}$, and $\mathrm{C_2H_2}$.
With this assumption, we derived the radius and density distributions of haze particles by the microphysical model.
Finally, we simulated transmission spectra of the atmospheres with the obtained profiles of haze particles and gaseous species.

For the temperature-pressure profile, we used the analytical model of \citet{Parmentier2014} with the default coefficient and opacity options \citep{Parmentier2015, Valencia2013}, in the same way as in section 5.2 of \citet{Kreidberg2022}.
As for the other atmospheric and haze parameters, we adopted the same values and references as those used in section 5.2 of \citet{Kreidberg2022}; solar elemental abundance ratios, eddy diffusion coefficient, and monomer radius, internal density, and refractive index (tholin and soot) of haze particles. 
We adopted a mass of 10 $M_{\earth}$.
For the UV spectrum of the host star, we used the spectrum of \citet{Duvvuri2023}, which was constructed based on the NICER X-ray and Hubble Space Telescope observations and empirical models. 
We calculated spectra with a range of $f_\mathrm{haze}$ from $10^{-5}$ to $1$ in 1 dex increments. The integrated monomer production rate for $f_\mathrm{haze} = 1$ (the sum of the photodissociation rates of $\mathrm{CH_4}$, $\mathrm{HCN}$, and $\mathrm{C_2H_2}$) is $2.41 \times 10^{-10}$ $\mathrm{g}$~$\mathrm{cm^2}$~$\mathrm{s}^{-1}$.

\section{Calculating interior temperature including tidal heating effects}  \label{appendix:tidal luminodity calc}

The total internal luminosity of a planet is a combination of residual heat of formation and any external sources such as tidal heating. The residual energy of formation is given by \citep[][]{rogers2010}:

\begin{equation} \label{eq:eq6}
    log \left(\frac{L_{int}}{L_{\odot}}\right)=a_{1}+a_{M}log \left(\frac{M_{p}}{M_{\oplus}}\right) + a_{R}log \left(\frac{R_{P}}{R_{J}}\right) + a_{t}log \left(\frac{t_{p}}{1 Gyr}\right)
\end{equation}

where $a_{1}$=-12.46 $\pm$ 0.05, $a_{M}$=1.74 $\pm$ 0.03, $a_{R}$=-0.94 $\pm$ 0.09 and $a_{t}$=-1.04 $\pm$ 0.04 where $t_{p}$ is the age of the planet in Gyr. The tidal luminosity depends on the eccentricity of the planet, tidal quality factor and planetary spin obliquity. The expressions for the calculation of $L_{tide}$ are given in Eqn 1-5 of \citet[][]{millholand_2020}. We assume 1$\%$ eccentricity based on ongoing TTV modeling efforts (private communication, John Livingston).

To estimate the internal temperature we assume the total luminosity due to residual energy of formation and tidal heating  to be dissipated due to blackbody emission:

\begin{equation} \label{eq:eq7}
   T_{int}=\left(\frac{L_{int}}{4\pi \sigma R_{p}^{2}}\right)^{1/4}
\end{equation}

\begin{figure}
    \centering
    \includegraphics[width=1\linewidth]{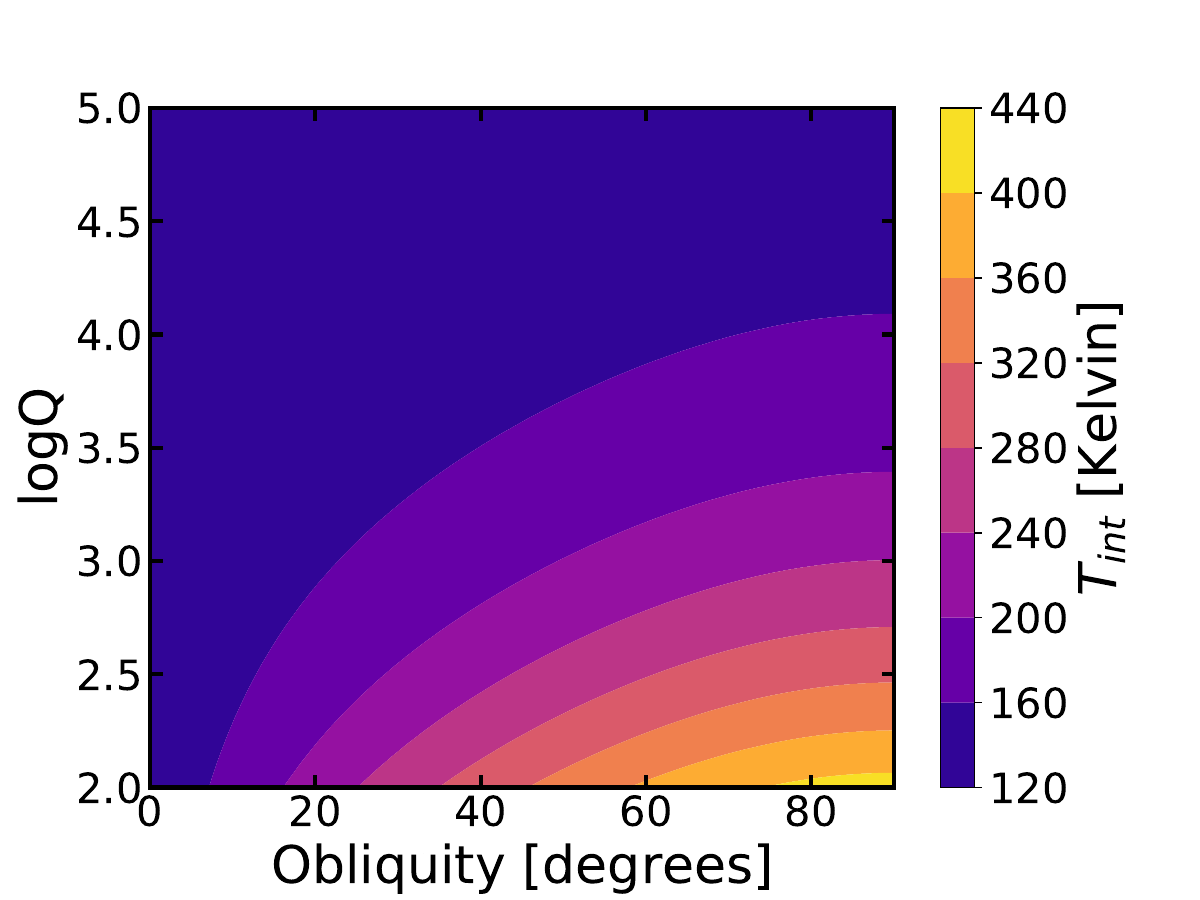}
    \caption{Internal temperature for V1298 Tau b for different values of tidal quality factor (logQ) and planetary spin axis obliquity. The tidal heating luminosity has been derived using formalism outlined in \citet{millholland2019,millholand_2020} (See Appendix \ref{appendix:tidal luminodity calc}). To estimate the total we add the energy of formation derived using scaling relations from \citet{rogers2010} to tidal luminosity. We convert the luminosoty to an internal temperature assuming a blackbody function and a planet radius equal to the radius of the radiative-convective boundary calculated in Figure \ref{fig:profile_comparison}. We assume 1$\%$ eccentricity based on ongoing TTV modelling efforts (private communication, John Livingston).  We require very high spin-orbit obliquities ($>80^{\circ}$) and very low tidal quality factor ($<$100) to reproduce internal temperatures higher than 400~K.  }
    \label{fig:external heating}
\end{figure}

We explore the parameter space of logQ-Obliquity and estimate the internal temperature for V1298 Tau b, which is shown in Figure \ref{fig:external heating}. We assume R$_{p}$ to be the planet radius at the radiative-convective boundary shown in Figure \ref{fig:profile_comparison}. We require high spin obliquity ($>$80$^{\circ}$) and tidal quality factor of the order of 100 to explain internal temperature higher than 400K. 

\end{document}